\documentclass[aps,prl,twocolumn,showpacs,superscriptaddress]{revtex4-2}
\usepackage{amsmath,amssymb}
\usepackage{bm}
\usepackage{graphicx}
\usepackage{booktabs}
\usepackage[dvipsnames]{xcolor}
\usepackage[colorlinks=true, linkcolor=blue, citecolor=blue, urlcolor=blue]{hyperref}
\newcommand{\smnote}[1]{{\color[gray]{0.35}(#1)}}
\newcommand{\prlsec}[1]{\par\vspace{0.5\baselineskip}\noindent{\color[RGB]{168,40,140}\textit{#1.}}---\ignorespaces}

\usepackage[normalem]{ulem}

\newcommand{\bea}{\begin{eqnarray}}
\newcommand{\eea}{\end{eqnarray}}

\newcommand{\la}{\langle}
\newcommand{\ra}{\rangle}

\newcommand{\tcr}[1]{\textcolor{red}{#1}}

\begin{document}

\title{Altermagnetic Ground States on the Ammann--Beenker Tiling}

\author{SeungBeom Hong}
\affiliation{Department of Physics, Hanyang University, Seoul 04763, Republic of Korea}
\author{YunHo Kim}
\affiliation{Department of Physics, Hanyang University, Seoul 04763, Republic of Korea}
\author{Moon Jip Park}
\email{moonjippark@hanyang.ac.kr}
\affiliation{Department of Physics, Hanyang University, Seoul 04763, Republic of Korea}
\affiliation{Hanyang Institute for Quantum Science and Quantum Technology, Hanyang University, Seoul 04763, Republic of Korea}
\affiliation{Research Institute for Natural Science and High Pressure, Hanyang University, Seoul, 04763, Republic of Korea}
\author{GiBaik Sim}
\email{gibaik.sim@unimelb.edu.au}
\affiliation{School of Physics, The University of Melbourne, Parkville, VIC 3010, Australia}
\date{\today}

\begin{abstract}
We identify exact altermagnetic ground states of a classical Heisenberg model on a quasicrystal. Considering the Ammann--Beenker tiling as an example, we determine the ground states by exploiting the parity structure of its parent four-dimensional hypercubic lattice. The ground state hosts coplanar altermagnetic order characterized by the preserved spin-group symmetry $[C_{2x}\parallel C_{8z}]$. Without translational symmetry, the momentum-space spin splitting that usually identifies altermagnets is not directly accessible. We instead couple the spins to itinerant electrons and show spin-resolved transport with $g$-wave symmetry, confirming the altermagnetic nature in a nonperiodic system. Our results establish a geometric route to altermagnetic order in quasicrystals, connecting exact magnetic ground states to their spin-resolved transport signatures.
\end{abstract}

\maketitle

\begin{figure*}
\centering
\includegraphics[width=1\linewidth]{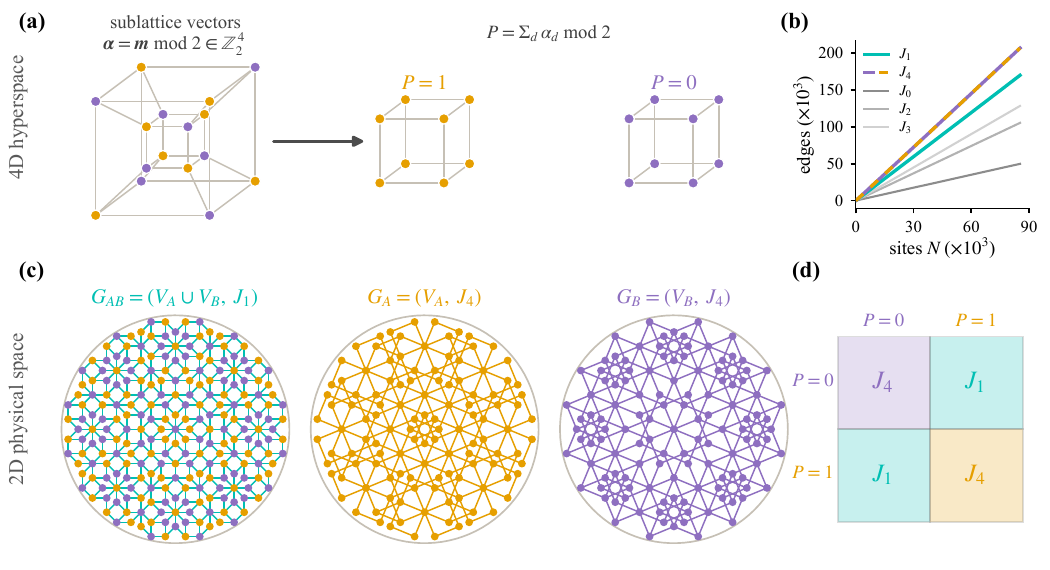}
\caption{{(a) Parity decomposition in the four-dimensional hypercubic parent lattice. The parity $P$ splits the $2^4$ sublattice vectors $\bm\alpha$ into two groups of eight, drawn as cubes for $V_A$ ($P=1$) and $V_B$ ($P=0$). (b) Number of bonds of each class $n$ versus the number of sites $N$. (c) Graphs in physical space after cut-and-project: $G_{AB}$ formed by the $J_1$ bonds, and $G_A$, $G_B$ formed by the $J_4$ bonds.}}
\label{fig1}
\end{figure*}

\prlsec{Introduction}Altermagnetism has recently attracted considerable attention as a magnetic order combining compensated magnetic moments with nonrelativistic spin splitting~\cite{Smejkal2022_PRX_Beyond,Smejkal2022_PRX040501,Mazin2022_PRX}. In altermagnets, opposite-spin sublattices are related by spatial rotations rather than by translation or inversion, allowing momentum-dependent spin splitting without net magnetization~\cite{Hayami2019_JPSJ,Yuan2020_PRB,Smejkal2022_PRX_Beyond}. This electronic structure enables unconventional transport responses, including the generation of spin-polarized currents through the spin-splitter effect~\cite{Gonzalez-Hernandez2021_PRL,Bai2022_PRL,Karube2022_PRL}. These properties make altermagnets promising platforms for spin transport without macroscopic magnetization.

While much of this research has focused on periodic crystals, recent studies have extended altermagnetism to nonperiodic systems, including amorphous materials and quasicrystals~\cite{dOrnellas2026_PRB,Chen2025_arXiv,Li2025_PRL}. Quasicrystals possess long-range structural order without translational periodicity and admit rotational symmetries forbidden in periodic crystals~\cite{Shechtman1984_PRL,Levine1984_PRL}. Their unconventional geometry shapes magnetic order and its spatial structure~\cite{Vedmedenko2004_PRL,ThiemChalker2015_PRB,KogaTsunetsugu2017_PRB,Koga2020_PRB,Koga2021_Socolar,KogaCoates2022_PRB,KogaMatsubara2025_PRB,Teixeira2025,Jeon2024_npjQM,Araujo2024_PRB}, while also influencing magnetic excitations, response, and critical behavior~\cite{Jeon2022_PRB_Magnon,Andrade2015_PRL,JeonSakai2026_arXiv,KogaSakai2025_PRB}. Studies of quasiperiodic and incommensurate systems have further revealed geometry-dependent long-range spin couplings and quantum correlations~\cite{Jeon2025_PRB_Hyperspace,Park2025_arXiv,Joung2025_arXiv}. Experimentally, long-range ferromagnetic and antiferromagnetic orders have been observed in icosahedral quasicrystals~\cite{Tamura2021_JACS,Tamura2025}.

Beyond supporting magnetic order, noncrystallographic rotational symmetries allow unconventional forms of altermagnetism, including $g$-, $h$-, and $i$-wave phases in octagonal, decagonal, and dodecagonal quasicrystals, respectively~\cite{Chen2025_arXiv,Li2025_PRL}. These phases were demonstrated through self-consistent mean-field treatments of interacting electronic models, with directional orbital degrees of freedom or decoration-induced hopping anisotropy enabling altermagnetic order~\cite{Chen2025_arXiv,Li2025_PRL}. The symmetry of magnetic order in quasicrystals has also been studied within the spin-space-group framework~\cite{Lifshitz1998_PRL}, with recent work developing a point-group classification of collinear order~\cite{Shao2026_npjCM}. Altermagnetic N\'eel states have also been demonstrated in bipartite quasicrystals using quantum Monte Carlo~\cite{Ji2025_arXiv}. These advances motivate a complementary question. \textit{Can altermagnetic order arise directly as the ground state of a spin model on a quasicrystalline tiling?}

In this work, we identify altermagnetic ground states of a classical Heisenberg model on the Ammann--Beenker (AB) tiling. The parity structure inherited from its four-dimensional (4D) parent lattice separates the tiling into two decoupled networks. One supports collinear altermagnetic order that breaks $C_{8z}$ but preserves $[\mathcal{T}\parallel C_{8z}]$, where $\mathcal{T}$ denotes time reversal. The other supports collinear N\'eel order that preserves $C_{8z}$. When coupled, the two networks form a single coplanar altermagnet with site-dependent canting. This state preserves the spin-group symmetry $[C_{2x}\parallel C_{8z}]$, which combines a $\pi$ spin rotation about the N\'eel ordering axis with an eightfold spatial rotation. Coupling the resulting spin texture to itinerant electrons, we demonstrate axis-selective spin transport with a $g$-wave angular dependence.

\prlsec{Spin model on the Ammann--Beenker tiling}The AB tiling is obtained by projecting the 4D hypercubic lattice $\mathbb{Z}^4$ onto a two-dimensional plane through the cut-and-project method~\cite{deBruijn1981_IM,DuneauKatz1985_PRL}. Each site of the tiling corresponds to a unique point $\bm{m}$ of the hypercubic lattice $\mathbb{Z}^4$. For instance, the positions of the sites in the two-dimensional physical plane are generated as
\begin{equation}
\bm{r}=\sum_{k=1}^{4}m_k\,\hat{\bm{u}}_k,
\label{eq:cut_project}
\end{equation}
where $\hat{\bm{u}}_k=(\cos\frac{(k-1)\pi}{4},\sin\frac{(k-1)\pi}{4})$ is the image of the unit vector $\bm{e}_k$ of the hypercubic lattice $\mathbb{Z}^4$. \smnote{The details of the cut-and-project method are given in Sec.~S1 of the Supplemental Material (SM)~\cite{SM}.}

The distances between the sites take discrete values, $r_0=\sqrt{2-\sqrt{2}}$, $r_1=1$, $r_2=\sqrt{2}$, $r_3=\sqrt{3}$, and $r_4=\sqrt{2+\sqrt{2}}$ in ascending order, in units of the tile edge. Among the corresponding bond classes $n=0,\ldots,4$, the tile edges ($n=1$) and the long diagonals of the rhombic tiles ($n=4$) are the two most frequent classes [Fig.~\ref{fig1}(b)]. Motivated by this classification, we consider a classical Heisenberg model with the exchange couplings $J_1$ and $J_4$ on these two bond classes,
\bea
H=J_1\sum_{\la i,j \ra_1}\bm{S}_i\cdot\bm{S}_j
+J_4\sum_{\la i,j \ra_4}\bm{S}_i\cdot\bm{S}_j,
\label{eq:H_spin}
\eea
where $\la i,j \ra_n$ denotes a pair of sites connected by a bond of class $n$.

It is important to note that, for our purpose, each site can be assigned parity numbers inherited from its 4D label $\bm{m}(i)$. We define two parity numbers of each site, $P$ and $Q$, as
\begin{equation}
P\equiv\alpha_1+\alpha_2+\alpha_3+\alpha_4,\quad Q\equiv\alpha_1+\alpha_3\ \ (\mathrm{mod}\ 2),
\label{eq:parity_def}
\end{equation}
where $\bm{\alpha}(i)\equiv\bm{m}(i)\ (\mathrm{mod}\ 2)\in\mathbb{Z}_2^4$ denotes the sublattice vector of site $i$ [Fig.~\ref{fig1}(a)]. The parity $P$ partitions the tiling into two sets of sites, $V_A=\{i\mid P(i)=1\}$ and $V_B=\{i\mid P(i)=0\}$. All tile edges are images of single unit vectors, $\Delta\bm{m}=\pm\bm{e}_k$. Each $J_1$ bond connects two sites whose sublattice vectors differ by
\begin{equation}
\Delta\bm{\alpha}_{J_1}\in\{(1000),(0100),(0010),(0001)\}.
\label{eq:dalpha_J1}
\end{equation}
Therefore, the two sites have opposite parities, $\Delta P=1$. Each $J_4$ bond corresponds to $\Delta\bm{m}=\pm(\bm{e}_k+\bm{e}_{k+1})$ with $\bm{e}_5\equiv-\bm{e}_1$. It connects two sites whose sublattice vectors differ in two cyclically adjacent components,
\begin{equation}
\Delta\bm{\alpha}_{J_4}\in\{(1100),(0110),(0011),(1001)\}.
\label{eq:dalpha_J4}
\end{equation}
Therefore, the two sites have the same parity, $\Delta P=0$. Each displacement also changes exactly one of $\alpha_1$ and $\alpha_3$. The two sites thus have opposite $Q$, $\Delta Q=1$. The $J_1$ bonds thus form a bipartite graph $G_{AB}$ between $V_A$ and $V_B$~\cite{diestel2024extremal}. The $J_4$ bonds split into two disconnected graphs, $G_A$ on $V_A$ and $G_B$ on $V_B$ [Fig.~\ref{fig1}(c)]. This bipartite structure allows us to determine the ground states analytically, as shown below.

\begin{figure*}
\centering
\includegraphics[width=1\linewidth]{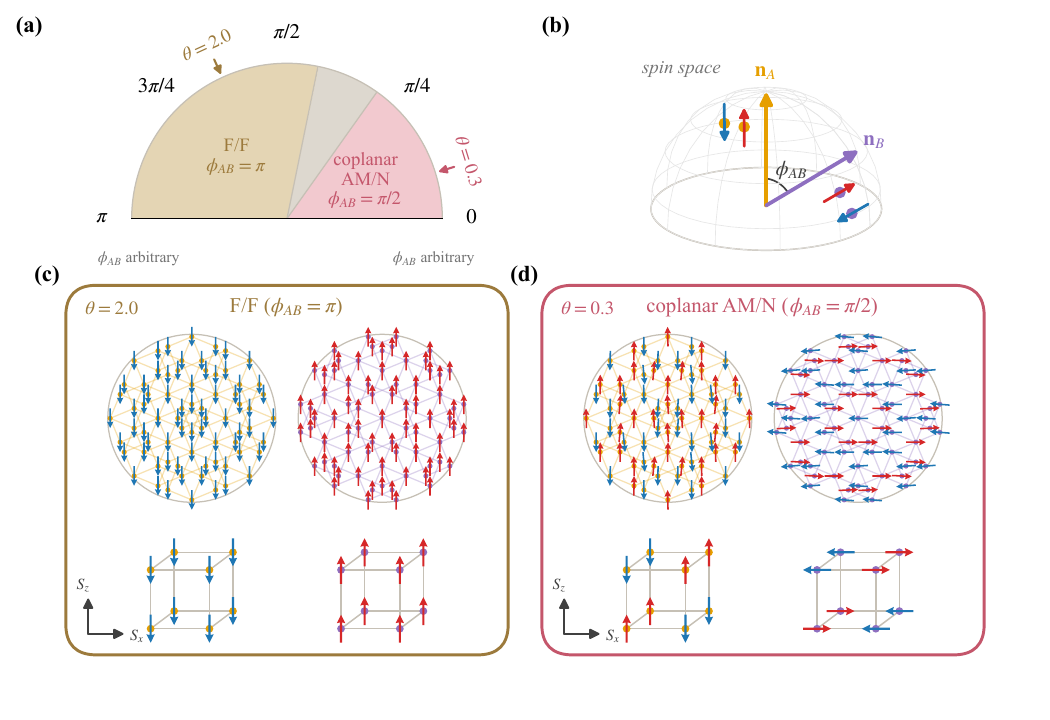}
\caption{(a) Ground-state phase diagram of the $J_1$--$J_4$ model for $\theta\in[0,\pi]$, with phase boundaries at $\theta\approx0.95$ and $1.37$. At $\theta=0$ and $\pi$, $G_A$ and $G_B$ decouple, having an arbitrary relative angle $\phi_{AB}$. (b) Relative angle $\phi_{AB}$ between the ordering axes $\mathbf{n}_A$ of $G_A$ and $\mathbf{n}_B$ of $G_B$ in spin space. (c) F/F and (d) coplanar AM/N ground states at the $\theta$ marked in (a): relaxed spins on $G_A$ and $G_B$, projected onto the $S_x$--$S_z$ plane (top), and their average over each sublattice $\bm\alpha$ on the cubes $V_A$ and $V_B$ (bottom).}
\label{fig2}
\end{figure*}

\prlsec{Phase diagram}To investigate the ground state of the $J_1$--$J_4$ model, we parametrize the couplings as $J_1=\sin\theta$ and $J_4=\cos\theta$ with $\theta\in[0,\pi]$. The cases $\theta=0$ and $\theta=\pi$ are the two analytically solvable points. Since $J_1=0$ at both points, the graphs $G_A$ and $G_B$ are completely decoupled. At $\theta=\pi$, the model reduces to a pure ferromagnetic $J_4$ model. In both $G_A$ and $G_B$, the $J_4$ bonds align the spins parallel to one another, yielding a collinear ferromagnetic ground state on each graph. A small antiferromagnetic $J_1$ away from this point aligns the two ferromagnetic graphs antiparallel.

On the other hand, at $\theta=0$, the model reduces to a pure antiferromagnetic $J_4$ model. Since $\Delta Q=1$ for every $J_4$ bond, each $J_4$ bond connects a site with $Q=0$ to a site with $Q=1$ within the same parity sector, $V_A$ or $V_B$. Each $J_4$ graph thus admits a bipartite coloring by $Q$, leaving no frustration. The ground state on each graph $G_\gamma$ is then
\begin{equation}
\bm{S}_i=(-1)^{Q(i)}\,\bm{n}_\gamma,
\label{eq:neel_Q}
\end{equation}
where the unit vector $\bm{n}_\gamma$ denotes the ordering axis of $G_\gamma$ ($\gamma=A,B$). This is a collinear N\'eel-type state on each graph. \smnote{The proof is given in Sec.~S2 of the SM.}

Interestingly, the two graphs respond differently to the eightfold rotation. The ground state on $G_A$ breaks $C_{8z}$ but preserves $[\mathcal{T}\parallel C_{8z}]$. The ground state on $G_B$ preserves $C_{8z}$. To see this, we compare a site $i$ with the site $\mathcal{R}i$ obtained by rotating it with $C_{8z}$. Since $C_{8z}$ maps $\bm{m}$ to $(-m_4,m_1,m_2,m_3)$, it acts on $\bm{\alpha}$ as the cyclic shift
\begin{equation}
C_{8z}:(\alpha_1,\alpha_2,\alpha_3,\alpha_4)\to(\alpha_4,\alpha_1,\alpha_2,\alpha_3).
\label{eq:cyclic_shift}
\end{equation}
The two parities thus transform as $P\to P$ and $Q\to Q+P$. Equation~\eqref{eq:neel_Q} then gives $\bm{S}_{\mathcal{R}i}=(-1)^{P(i)}\bm{S}_i$. On $G_A$ with $P=1$, each site carries a spin opposite to that of its $C_{8z}$ partner. This state is thus invariant under $[\mathcal{T}\parallel C_{8z}]$ but not under $C_{8z}$, which is precisely the symmetry of a collinear altermagnet. On $G_B$ with $P=0$, each site carries the same spin as its $C_{8z}$ partner. This state is thus a collinear N\'eel state that preserves $C_{8z}$ [cf.~Fig.~\ref{fig2}(d)]. The relative angle $\phi_{AB}$ between $\bm{n}_A$ and $\bm{n}_B$ remains arbitrary at this point. A small $J_1$ lifts this degeneracy by locking $\phi_{AB}$ at $\pi/2$, as shown below.

Having established these two limits, we now turn to the numerical phase diagram for $\theta\in[0,\pi]$, shown in Fig.~\ref{fig2}(a). We obtain the ground states on $C_{8z}$-symmetric AB tilings containing up to $10^4$ sites by iterative energy minimization. Starting from random spin configurations, we repeatedly relax each spin toward its local energy-minimizing direction for up to $5\times10^{4}$ iterations $(3\times 10^{5}$ near the phase boundaries), and keep the lowest-energy state.

For $0\leq\theta\lesssim0.95$, $G_A$ exhibits collinear altermagnetic order, whereas $G_B$ exhibits collinear N\'eel order. We refer to this regime as the altermagnet/N\'eel (AM/N) phase. As shown below, the two networks together form a coplanar altermagnet. For finite $\theta$ within the AM/N phase, the antiferromagnetic $J_1$ locks the two ordering axes orthogonal, $\phi_{AB}=\pi/2$ [Fig.~\ref{fig2}(b)]. This locking follows from the parity structure. Since $C_{8z}$ reverses the spins on $G_A$ but not on $G_B$, the $J_1$ bonds cost no energy for any $\phi_{AB}$ as long as the spins stay collinear. The local $J_1$ fields thus cancel on average. They do not vanish site by site, however. Their components perpendicular to each ordering axis cant the spins within the plane spanned by $\bm{n}_A$ and $\bm{n}_B$ [Fig.~\ref{fig2}(d)]. The resulting energy gain is maximal for orthogonal axes. Taking $\bm{n}_B$ as the $x$ axis, the canted state preserves the spin-group symmetry $[C_{2x}\parallel C_{8z}]$, a $\pi$ spin rotation about $\bm{n}_B$ combined with the eightfold rotation. This symmetry exists only for orthogonal axes, since the spin rotation must leave $\bm{n}_B$ invariant while reversing $\bm{n}_A$. For $0.95\lesssim\theta\lesssim1.37$, energy minimization from random initial states does not converge to a unique ordered state.

At $\theta\approx1.37$, the system enters the ferromagnet/ferromagnet (F/F) phase. Unlike the canted AM/N phase, the F/F phase is fully collinear: each graph is ferromagnetically ordered, and their magnetizations are antiparallel, with $\phi_{AB}=\pi$ [Fig.~\ref{fig2}(c)]. This phase persists up to $\theta=\pi$, where $J_1$ vanishes and $\phi_{AB}$ again becomes arbitrary. Below, we focus on this canted AM/N phase. \smnote{Numerical methods, order parameters, stability analysis, and the perturbative treatment of the canting are given in Secs.~S3 and S4 of the SM.}

\prlsec{Transport signature of altermagnetism}To connect the spin texture to electronic states, we couple the spins to itinerant electrons through an onsite exchange coupling. The electrons hop along the bonds of class $n=1$,
\begin{equation}
H_{\rm el}=t\sum_{\la i,j\ra_1}\bigl(c_i^\dagger c_j+\mathrm{H.c.}\bigr)
+h\sum_i c_i^\dagger\,(\bm{m}_i\cdot\bm{\sigma})\,c_i,
\label{eq:Hel}
\end{equation}
where $c_i=(c_{i\uparrow},c_{i\downarrow})^T$ annihilates an electron at site $i$, $\bm{m}_i$ denotes the spin at site $i$ in the coplanar AM/N spin texture, $\bm{\sigma}=(\sigma_x,\sigma_y,\sigma_z)$ denotes the Pauli matrices in spin space, and $h$ is the exchange coupling between the itinerant electrons and the spins. Below we set the hopping amplitude $t=1$. Without loss of generality, we choose the ordering axes on $G_A$ and $G_B$ along the $z$ and $x$ directions, respectively.

\begin{figure}
\centering
\includegraphics[width=1\linewidth]{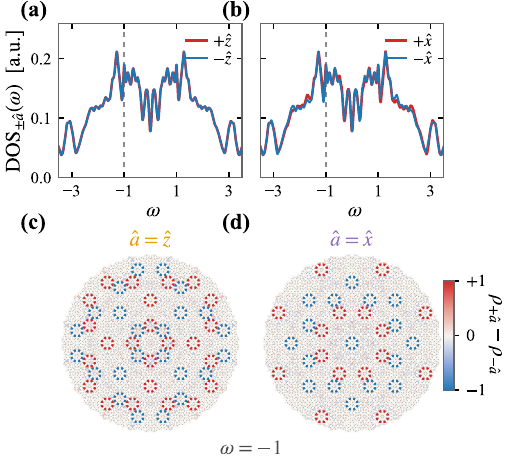}
\caption{Spin-resolved electronic structure of the canted AM/N state at $h=1$, $\theta=0.3$, and $\eta=0.05$. (a) DOS resolved along $\hat z$, the ordering axis of $G_A$, whose two channels coincide by $[C_{2x}\parallel C_{8z}]$. (b) DOS resolved along $\hat x$, the ordering axis of $G_B$, whose two channels split. (c),(d) LDOS spin splitting $\rho_{+\hat a}-\rho_{-\hat a}$ at $\omega=-1$, normalized to its maximum, for (c) $\hat a=\hat z$ and (d) $\hat a=\hat x$.}
\label{fig4}
\end{figure}

We characterize the spin polarized local density of states (LDOS)~\cite{Sukhachov2024} along a polarization axis
$\hat a\in\{\hat{x},\hat{y},\hat{z}\}$,
\begin{equation}
\rho_{\pm\hat{a}}(\bm{r},\omega)=-\frac{1}{\pi}\sum_{n}\mathrm{Im}\,
\frac{\psi_{n}^{\dagger}(\bm{r})\,\mathcal{P}_{\pm\hat{a}}\,\psi_n(\bm{r})}{\omega-E_n+i\eta},
\label{eq:ldos}
\end{equation}
where $\mathcal{P}_{\pm\hat{a}}=(\sigma_0\pm\hat{a}\cdot\bm{\sigma})/2$ projects onto spin $\pm\hat{a}$. Here, $E_n$ and $\psi_n$ are the eigenvalues and eigenvectors of the Hamiltonian $H_{\rm el}$, $\eta$ is a Lorentzian broadening, and $\omega$ is the energy of probing. The corresponding density of states (DOS) is $\mathrm{DOS}_{\pm\hat{a}}(\omega)=\sum_{\bm{r}}\rho_{\pm\hat{a}}(\bm{r},\omega)$. For $h=0$, the hopping Hamiltonian on the AB tiling hosts an extensively degenerate manifold of strictly confined zero-energy states~\cite{KohmotoSutherland1986_PRL,Arai1988_PRB,Rieth1998_JPCM,DayRoberts2020_PRB,PhysRevB.105.045146,Keskiner2022_PRB,Koga2020_PRB}. A finite exchange coupling lifts this degeneracy and gathers a subset of these states into narrow levels near $\omega=\pm h$. We therefore probe the spin texture at $\omega=-h$, where these confined levels appear as octagonal rings in the spin-resolved LDOS [Figs.~\ref{fig4}(c) and \ref{fig4}(d)].

We now examine how the spin-group symmetry $[C_{2x}\parallel C_{8z}]$ of the AM/N phase is encoded in the DOS. Since $C_{2x}$ reverses $S_z$, the $\hat{z}$-resolved LDOS splitting changes sign under $C_{8z}$, with zero net magnetization. This realizes the altermagnetic spin splitting [Fig.~\ref{fig4}(c)]. By contrast, since $C_{2x}$ leaves $S_x$ invariant, the $\hat{x}$-resolved splitting is invariant under $C_{8z}$ [Fig.~\ref{fig4}(d)]. The symmetry therefore allows a residual polarization along $\hat{x}$, which splits the two $\hat{x}$ channels of the DOS with a sign and magnitude that depend on the tiling [Fig.~\ref{fig4}(b)] and has no counterpart along $\hat{z}$. Below, we show that the same axis selectivity governs the spin-resolved transmission.

\begin{figure}
\centering
\includegraphics[width=1\linewidth]{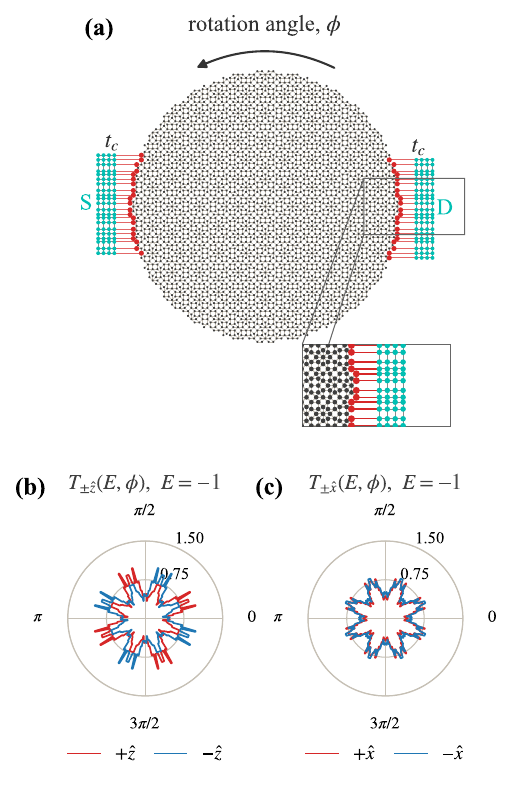}
\caption{Spin-resolved transmission through the canted AM/N state at $\theta=0.3$, $h=1$ and $E=-1$, plotted radially versus rotation angle $\phi$. (a) Device geometry: nonmagnetic square-lattice leads, source $S$ and drain $D$, coupled by $t_c$ to the tiling and rotated jointly by $\phi$, with the contact region enlarged in the inset. (b) Transmission $T_{\pm\hat z}$ resolved along $\hat z$, the ordering axis of $G_A$, whose two channels are exchanged by $[C_{2x}\parallel C_{8z}]$ under $\phi\to\phi+\pi/4$, $T_{+\hat z}(\phi+\pi/4)=T_{-\hat z}(\phi)$. (c) Transmission $T_{\pm\hat x}$ resolved along $\hat x$, the ordering axis of $G_B$, whose two channels each preserve $C_{8z}$.}
\label{fig5}
\end{figure}

To probe this axis selectivity in transport, we attach two nonmagnetic semi-infinite square-lattice leads~\cite{Lewenkopf_2013,Hallberg2025_PRB}, a source $S$ and a drain $D$, to the outermost sites of the tiling with contact hopping $t_c=1$, and rotate the contacts jointly by $\phi$ [Fig.~\ref{fig5}(a)]. For each axis $\hat a\in\{\hat x,\hat z\}$, the source injects electrons with spin either $+\hat a$ or $-\hat a$, and the drain collects electrons of both spins. The resulting transmission~\cite{Datta1995} at the energy $E=-h$ of the confined levels is defined by
\bea
T_{\pm\hat{a}}(E,\phi)=\mathrm{Tr}\bigl[\Gamma_D G\,\Gamma_S^{\pm\hat{a}}G^\dagger\bigr],
\label{eq:T}
\eea
where $G=[E+i0^+-H_{\rm el}-\Sigma_S-\Sigma_D]^{-1}$ is the retarded Green's function of the device and $\Gamma_{S,D}=i(\Sigma_{S,D}-\Sigma_{S,D}^\dagger)$ are the lead broadening matrices, with $\Gamma_S^{\pm\hat a}=\Gamma_S\,\mathcal{P}_{\pm\hat a}$ the source broadening projected onto spin $\pm\hat a$, while $\Gamma_D$ remains unprojected. The leads enter only through their self-energies $\Sigma_{S,D}$, which are nonzero only on the sites in contact with the source and the drain, so that $\phi$ moves the contact positions while $H_{\rm el}$ stays fixed. Since the tiling and the contacts are $C_{8z}$ symmetric, rotating the contacts by $\pi/4$ is equivalent to rotating the spin texture by $C_{8z}$. Because $C_{2x}$ reverses $S_z$, this rotation exchanges the two spin channels along $\hat z$,
\bea
T_{+\hat{z}}(\phi+\pi/4)=T_{-\hat{z}}(\phi),
\label{eq:C8T}
\eea
as shown in Fig.~\ref{fig5}(b). The spin-up and spin-down transmissions along $\hat z$ alternate under every $\pi/4$ rotation, realizing an altermagnetic spin splitting that is odd under $\phi\to\phi+\pi/4$, i.e., of $g$-wave symmetry. By contrast, since $C_{2x}$ leaves $S_x$ invariant, each $T_{\pm\hat{x}}$ is $C_{8z}$ symmetric and shows no such alternation [Fig.~\ref{fig5}(c)]. Their difference $T_{+\hat{x}}-T_{-\hat{x}}$ is therefore allowed by the symmetry, while its sign and magnitude depend on the contact position and the tiling. The transmission thus inherits the axis selectivity of the DOS: resolved along the ordering axis of $G_A$, it exhibits the altermagnetic alternation, whereas along that of $G_B$ it remains $C_{8z}$ symmetric. \smnote{The lead model and additional transport data are given in Sec.~S5 of the SM.}

\prlsec{Conclusion}We have established altermagnetic ground states of a classical Heisenberg model on the AB tiling. We traced their origin to the parity structure of the 4D parent lattice. Competing exchanges select orthogonal ordering axes through coplanar canting while preserving the spin-group symmetry $[C_{2x}\parallel C_{8z}]$. The resulting spin texture yields axis-selective spin transport with a $g$-wave angular dependence. While we have focused on the $J_1$--$J_4$ model, future work could determine the stability of this phase against additional exchanges, such as $J_2$ and $J_3$. Extending the model to quantum spins would further clarify how quantum fluctuations affect the altermagnetic order and the orthogonal locking of the two ordering axes.

\prlsec{Acknowledgments}We thank Jemin Park, SungBin Lee, and Junmo Jeon for bringing their work to our attention. G.B.S. also thanks Stephan Rachel and Peru d'Ornellas for fruitful discussions related to this work. This work was supported by the National Research Foundation of Korea (NRF) grant funded by the Korea government (MSIT) (Grants No. RS-2025-16070482, RS-2025-25446099, RS\tcr{-}2025-03392969, RS-2025-25464760, RS-2025-25431194, RS-2025-25433233). G.B.S. acknowledges support from the Australian Research Council (ARC) through Grant No. DP240100168 and the NRF through Grant No. RS-2024-00453943. This research was undertaken using resources from the National Computational Infrastructure (NCI Australia), an NCRIS enabled capability supported by the Australian Government.

\bibliography{references}
\end{document}


\title{Supplemental Material for\\ ``Altermagnetic Ground States on the Ammann--Beenker Tiling''}
\maketitle
\tableofcontents

In Sec.~\ref{sec:setup}, we construct the tiling by four-dimensional (4D) cut-and-project and define the sublattice and parity labels of each site, together with the action of the eightfold rotation $C_{8z}$ in 4D. These labels fix, in Sec.~\ref{sec:J4_parity}, the staggered sign of the $J_4$ order and the parity of the $J_4$ paths between $C_{8z}$ partners. Building on the staggered sign, we define in Sec.~\ref{sec:order_parameters} the order parameters of the altermagnet/N\'eel (AM/N) and ferromagnet/ferromagnet (F/F) phases and locate the phase boundaries of the ground-state phase diagram, which are largely insensitive to the patch radius $R$. In Sec.~\ref{sec:canting}, we show that a small $J_1$ locks the two ordering axes orthogonally through a second-order canting, realizing the canted AM/N state. Finally, in Sec.~\ref{sec:transport}, we specify the device, the leads, and the angle-dependent contacts behind the transmission and spin polarization of the main text.

\section{Geometric framework: four-dimensional cut-and-project, parity, and rotation}
\label{sec:setup}

The Ammann--Beenker tiling is a two-dimensional aperiodic structure with eightfold rotational symmetry obtained by cut-and-project from the 4D integer lattice. We use circular patches of cutoff radius $R$, with $N\sim R^2$ sites, and denote the physical coordinates of site $i$ by $\br(i)=(x_i,y_i)$.
The distinguished coupling is the parity-preserving coupling $J_4$ at bond length $\sqrt{2+\sqrt{2}}\approx1.848$, which partitions the lattice into two disjoint graphs, $G_B$ and $G_A$, with vertex sets $V_B$ and $V_A$; we place the origin $\br=(0,0)$ in $V_B$.

\subsection{Four-dimensional cut-and-project}
\label{sec:cap}

The physical and perpendicular projection matrices are
\bea
A = \begin{pmatrix} 1 & 0 \\ 1/\sqrt{2} & 1/\sqrt{2} \\ 0 & 1 \\ -1/\sqrt{2} & 1/\sqrt{2} \end{pmatrix},
\qquad
A_\perp = \begin{pmatrix} 1 & 0 \\ -1/\sqrt{2} & 1/\sqrt{2} \\ 0 & -1 \\ 1/\sqrt{2} & 1/\sqrt{2} \end{pmatrix},
\label{eq:Amat}
\eea
so that an integer 4-vector $\bmm\in\ZZ^4$ has physical and perpendicular images $\br=\bmm A$ and $\br_\perp=\bmm A_\perp$. The $k$th row of $A$ is the unit vector $\hat{\bm u}_k=(\cos\frac{(k-1)\pi}{4},\sin\frac{(k-1)\pi}{4})$ of Eq.~(1) of the main text, so that $\br=\sum_k m_k\hat{\bm u}_k$. A 4D lattice point $\bmm$ is accepted as a site of the Ammann--Beenker tiling if and only if its perpendicular-space image $\br_\perp(\bmm)$ lies inside the acceptance window $\mathcal{W}$: the regular octagon whose vertices are
\bea
\tfrac{1}{2}(\pm a_0,\,\pm 1)\quad\text{and}\quad\tfrac{1}{2}(\pm 1,\,\pm a_0),\qquad a_0=1+\sqrt{2},
\label{eq:window}
\eea
a region fixed independently of $N_{4D}$. For finite approximants of (even) size $N_{4D}$ we take $\bmm\in\{-N_{4D}/2,\ldots,N_{4D}/2\}^4$ with an additional radial cutoff $|\br|<R$.

\subsection{Sublattice vectors and parities}
\label{sec:labels}
For each site $i$ we define
\bea
\balpha(i) &\equiv& \bmm(i)\ (\mathrm{mod}\ 2) \in \ZZ_2^4, \label{eq:alpha} \\
P(i) &\equiv& \alpha_1+\alpha_2+\alpha_3+\alpha_4,\qquad Q(i)\equiv\alpha_1+\alpha_3\ \ (\mathrm{mod}\ 2), \label{eq:parity}
\eea
the sublattice vector $\balpha$ and the two parities $P$ and $Q$ [Eq.~(3) of the main text]. The parity $P$ partitions the sites into $V_A=\{i\mid P(i)=1\}$ and $V_B=\{i\mid P(i)=0\}$.

\subsection{Four-dimensional parity and the $J_4$ bipartition}
\label{sec:parity_J4}
For $R\in\{20,80\}$ the bipartite class assigned by the $J_4$ graph coincides exactly with the 4D parity (Table~\ref{tab:cross_parity}): the count $N_{P,\gamma}=\#\{i:P(i)=P,\gamma(i)=\gamma\}$ is diagonal under $P\in\{1,0\}\Leftrightarrow\gamma\in\{V_A,V_B\}$, with all off-diagonal entries vanishing.

\begin{table}[h]
\begin{tabular}{ccccc}
\toprule
& \multicolumn{2}{c}{$R=20$ ($N=1521$)} & \multicolumn{2}{c}{$R=80$ ($N=24289$)} \\
\cmidrule(lr){2-3}\cmidrule(lr){4-5}
& $\gamma=B$ & $\gamma=A$ & $\gamma=B$ & $\gamma=A$ \\
\midrule
$P=0$ & 753 & 0 & 12185 & 0 \\
$P=1$ & 0 & 768 & 0 & 12104 \\
\bottomrule
\end{tabular}
\caption{Numerical confirmation that the $J_4$ bipartition coincides with the 4D parity. Entries are the joint counts $N_{P,\gamma}=\#\{i:P(i)=P,\ \gamma(i)=\gamma\}$ of sites with parity $P$ [Eq.~\eqref{eq:parity}] and $J_4$-graph vertex set $\gamma\in\{V_A,V_B\}$, for two cutoff radii $R=20$ ($N=1521$) and $R=80$ ($N=24289$). The matrix is exactly diagonal under $P=1\Leftrightarrow V_A$ and $P=0\Leftrightarrow V_B$, with all off-diagonal entries vanishing, so the two labelings are identical.}
\label{tab:cross_parity}
\end{table}

The geometric origin is direct. The $J_4$ distance corresponds in 4D to the displacements $\Delta\bmm$ related to $(1,1,0,0)$ by $C_{8z}$. Projecting the representative with $A$,
\bea
\br\bigl((1,1,0,0)\bigr) = (1,1,0,0)\,A = \bm{e}_1A+\bm{e}_2A = \bigl(1+\tfrac{1}{\sqrt{2}},\ \tfrac{1}{\sqrt{2}}\bigr),
\label{eq:r_J4_proj}
\eea
where $\bm{e}_kA$, the $k$th row of $A$, is the physical image of the unit vector $\bm{e}_k$ of $\ZZ^4$, gives $|\br|^2=(1+\tfrac{1}{\sqrt2})^2+\tfrac12=2+\sqrt2$, confirming the $J_4$ bond length $\sqrt{2+\sqrt2}$. Since $(1,1,0,0)$ has $\Delta P=0$, $J_4$ edges preserve parity and connect only equal-parity sites; the $J_4$ graph therefore disconnects into the graphs $G_A$ and $G_B$, each containing sites of a single parity.
Table~\ref{tab:couplings} summarizes the parity character of the two principal couplings.

\begin{table}[h]
\begin{tabular}{ccccc}
\toprule
coupling & bond length & repr.\ $\Delta\bmm$ & $\Delta P$ & action \\
\midrule
$J_1$ & $1$ & $(1,0,0,0)$ & 1 & forms $G_{AB}$ between $V_A$ and $V_B$ \\
$J_4$ & $\sqrt{2+\sqrt{2}}$ & $(1,1,0,0)$ & 0 & within $G_A$ or $G_B$ \\
\bottomrule
\end{tabular}
\caption{
Parity character of the two principal exchange couplings. For each coupling we list the physical-plane bond length, a representative 4D displacement $\Delta\bmm$ (Sec.~\ref{sec:parity_J4}), and the induced parity change $\Delta P=\sum_k\Delta m_k\bmod 2$. The parity-flipping $J_1$ has $\Delta P=1$ and connects the two sets, whereas the parity-preserving $J_4$ has $\Delta P=0$ and acts only within a set, disconnecting the $J_4$ graph into $G_A$ and $G_B$.}
\label{tab:couplings}
\end{table}

\subsection{Eightfold rotation in four dimensions}
\label{sec:rotation}
The physical-plane rotation $C_{8z}$ is represented by an integer matrix $M$ that maps the 4D lattice onto itself, obeying $MA=A\,C_{8z}$, which fixes
\bea
M=\begin{pmatrix} 0 & 0 & 0 & -1\\ 1 & 0 & 0 & 0\\ 0 & 1 & 0 & 0\\ 0 & 0 & 1 & 0 \end{pmatrix},
\qquad M\bm{m} = (-m_4,m_1,m_2,m_3)^{T},
\label{eq:M}
\eea
with $M^8=I$, $M^4=-I$. Reducing modulo 2 turns the sign into $+1$ ($-1\equiv1$), so on the space $\ZZ_2^4$ of sublattice vectors the rotation collapses to the order-four cyclic shift
\bea
C_{8z}:(\alpha_1,\alpha_2,\alpha_3,\alpha_4)\to(\alpha_4,\alpha_1,\alpha_2,\alpha_3).
\label{eq:cyclic}
\eea
Thus $C_{8z}$ acts with order four on the sublattice vectors, and $C_{2z}=C_{8z}^4$ acts trivially.

Hence, the eight $C_{8z}$ images of a site remain in its set. The shift also maps $\alpha_1+\alpha_3$ to $\alpha_2+\alpha_4$. The second parity $Q\equiv\alpha_1+\alpha_3$ [Eq.~(3) of the main text] thus transforms as $Q\to Q+P$. Under $C_{8z}$, the $2^4$ sublattice vectors fall into six orbits (Table~\ref{tab:orbits}).

\begin{table}[h]
\centering
\begin{tabular}{ccc}
	\toprule
	orbit & set & $\bm{\alpha}$ \\
	\midrule
	1 & $V_A$ $(P=1)$ & $(1000),\;(0100),\;(0010),\;(0001)$ \\
	2 & $V_A$ $(P=1)$ & $(1110),\;(0111),\;(1011),\;(1101)$ \\
	\midrule
	3 & $V_B$ $(P=0)$ & $(0000)$ \\
	4 & $V_B$ $(P=0)$ & $(1111)$ \\
	5 & $V_B$ $(P=0)$ & $(1010),\;(0101)$ \\
	6 & $V_B$ $(P=0)$ & $(1100),\;(0110),\;(0011),\;(1001)$ \\
	\bottomrule
\end{tabular}
\caption{The sublattice vectors $\bm{\alpha}\in\mathbb{Z}_2^4$ are grouped into six distinct orbits under $C_{8z}$. Within each orbit, the vectors are ordered such that each successive application of $C_{8z}$ maps one vector to the next, with the last vector mapped back to the first.}
\label{tab:orbits}
\end{table}

\section{Parity of the \texorpdfstring{$J_4$}{J4} paths between \texorpdfstring{$C_{8z}$}{C8z} partners}
\label{sec:J4_parity}

\subsection{staggered sign}
\label{sec:neel_sign}
At $\theta=0$, the ground state on each graph $G_\gamma$ ($\gamma=A,B$) reverses the spin across every antiferromagnetic $J_4$ bond. We encode this reversal by the staggered sign
\bea
\epsilon_i=(-1)^{Q(i)},\qquad Q\equiv\alpha_1+\alpha_3\ (\mathrm{mod}\ 2),
\label{eq:neel_sign}
\eea
where $\alpha_k(i)$ denotes the $k$th component of the sublattice vector $\balpha(i)$ of site $i$. Each of the four $J_4$ displacements $\Delta\balpha\in\{(1100),(0110),(0011),(1001)\}$ in Eq.~(5) of the main text changes exactly one of $\alpha_1$ and $\alpha_3$, so $\epsilon_i\epsilon_j=-1$ for every $J_4$ bond $\langle i,j\rangle_4$. Hence, the collinear state $\bS_i=\epsilon_i\bm n_{\gamma(i)}$ satisfies every $J_4$ bond and reaches the lower bound $-J_4$ per bond, realizing the N\'eel ground state of the pure $J_4$ model. Here, $\bm n_\gamma$ denotes the ordering axis of $G_\gamma$, and $\gamma(i)$ the graph containing site $i$. The staggered sign also defines the N\'eel vectors and the AM/N order parameter in Sec.~\ref{sec:order_parameters}.

\subsection{Action of \texorpdfstring{$C_{8z}$}{C8z} on the staggered sign}
\label{sec:neel_rotation}
Having established the staggered sign, we now examine how it transforms under $C_{8z}$. We denote by $\mathcal{R}$ the permutation of the sites induced by $C_{8z}$, $i\mapsto\mathcal{R}i$. Since $C_{8z}$ acts on the sublattice vector as the cyclic shift $(\alpha_1,\alpha_2,\alpha_3,\alpha_4)\to(\alpha_4,\alpha_1,\alpha_2,\alpha_3)$ [Eq.~\eqref{eq:cyclic}], the rotated site has $\epsilon_{\mathcal{R}i}=(-1)^{\alpha_2(i)+\alpha_4(i)}$. Multiplying by $\epsilon_i$ gives
\bea
\epsilon_i\,\epsilon_{\mathcal{R}i}=(-1)^{\alpha_1(i)+\alpha_2(i)+\alpha_3(i)+\alpha_4(i)}=(-1)^{P(i)}.
\label{eq:eps_rotation}
\eea
Therefore, $C_{8z}$ reverses the staggered sign at every site of $G_A$ ($P=1$) and preserves it at every site of $G_B$ ($P=0$), independently of the sublattice vector.

\subsection{Parity of the \texorpdfstring{$J_4$}{J4} paths}
\label{sec:path_parity}
Equation~\eqref{eq:eps_rotation} fixes the parity of every $J_4$ path between a site and its rotated image. Consider a path of $J_4$ bonds connecting $i$ and $\mathcal{R}i$, and let $\ell$ denote the number of bonds along it. The path stays within one graph because $J_4$ bonds preserve $P$. Since the staggered sign reverses along each bond, $\epsilon_i\epsilon_{\mathcal{R}i}=(-1)^\ell$, and Eq.~\eqref{eq:eps_rotation} gives
\bea
\ell\equiv P(i)\pmod 2.
\label{eq:path_parity}
\eea
Hence, every $J_4$ path from $i$ to $\mathcal{R}i$ has odd length in $G_A$ and even length in $G_B$, independently of the route. Such a path exists because $G_A$ and $G_B$ are each connected on the $C_{8z}$-symmetric patches used here. The same connectivity makes $\bS_i=\epsilon_i\bm n_{\gamma(i)}$ the unique ground state of each graph up to a global spin rotation.

For this ground state, Eq.~\eqref{eq:eps_rotation} gives $\bS_{\mathcal{R}i}=-\bS_i$ on $G_A$ and $\bS_{\mathcal{R}i}=\bS_i$ on $G_B$. The N\'eel state of $G_A$ thus breaks $C_{8z}$ but preserves $\CT$, realizing a collinear altermagnet, whereas the N\'eel state of $G_B$ preserves $C_{8z}$. Equation~\eqref{eq:eps_rotation} also underlies the cancellation of the first-order $J_1$ energy and the spin-group symmetry of the canted state in Sec.~\ref{sec:canting}.

\section{Order parameters and phase diagram}
\label{sec:order_parameters}

\subsection{Order parameters}
\label{sec:op_def}
Each set is itself bipartite under $J_4$, with the sites of $\epsilon_i=+1$ and $\epsilon_i=-1$ [Eq.~\eqref{eq:neel_sign}] as its two sublattices. We define the N\'eel vector of $G_\gamma$ as the difference of the two sublattice magnetizations,
\bea
\bm M_\gamma\equiv\sum_{i\in V_\gamma}\epsilon_i\bm S_i=\sum_{i\in V_\gamma,\ \epsilon_i=+1}\bm S_i-\sum_{i\in V_\gamma,\ \epsilon_i=-1}\bm S_i ,\qquad \bm n_\gamma\equiv\bm M_\gamma/|\bm M_\gamma| ,
\label{eq:neel}
\eea
with $\bm M_\gamma$ the staggered magnetization and $\bm n_\gamma$ its ordering axis. Here, the relative angle $\phi_{AB}$ between the ordering axes of $G_A$ and $G_B$ is given by $\cos\phi_{AB}=\bm n_A\cdot\bm n_B$. With the magnetization $\bm m_\gamma\equiv\sum_{i\in V_\gamma}\bS_i$ of each set, we define the order parameters of the AM/N, F/F, and ferromagnetic (FM) phases as
\bea
\Lambda_{\rm AM/N}=\frac{2}{N^2}\bigl(|\bm M_A|^2+|\bm M_B|^2\bigr),\qquad
\Lambda_{\rm F/F}=\frac{1}{N^2}|\bm m_A-\bm m_B|^2,\qquad
\Lambda_{\rm FM}=\frac{1}{N^2}|\bm m_A+\bm m_B|^2,
\label{eq:Lambda}
\eea
where $N$ denotes the number of sites. The AM/N order parameter measures the N\'eel amplitude of both sets independently of $\phi_{AB}$, whereas $\phi_{AB}$ gives their relative orientation. Likewise, $\Lambda_{\rm F/F}$ measures two oppositely polarized ferromagnetic orders on $G_A$ and $G_B$, whereas $\Lambda_{\rm FM}$ measures the uniform ferromagnetic order. Each quantity is close to unity in its own collinear state and vanishes in the other.

\subsection{Phase diagram}
\label{sec:phase_diagram}
\begin{figure}[h]
\centering
\includegraphics[width=\linewidth]{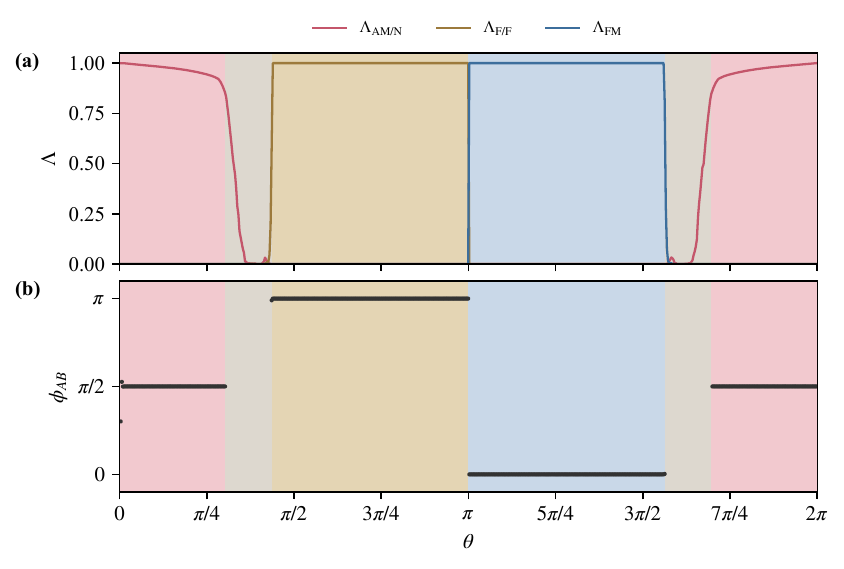}
\caption{(a) Order parameters $\Lambda_{\rm AM/N}$, $\Lambda_{\rm F/F}$, and $\Lambda_{\rm FM}$ [Eq.~\eqref{eq:Lambda}] of the ground state versus $\theta\in[0,2\pi]$ ($N=3409$). The background color marks the AM/N, noncoplanar, F/F, FM, noncoplanar, and AM/N phases, separated at $\theta=0.953$, $1.37$, $\pi$, $4.915$, and $2\pi-0.953$. (b) Relative angle $\phi_{AB}$ between the ordering axes $\bm n_A$ and $\bm n_B$ of $G_A$ and $G_B$: $\pi/2$ in the AM/N phase, $\pi$ in the F/F phase, and $0$ in the FM phase. It is not shown in the noncoplanar phase or at $\theta=0,\pi,2\pi$, where it is arbitrary.}
\label{fig:phase_order}
\end{figure}

We compute the order parameters $\Lambda_{\rm AM/N}$, $\Lambda_{\rm F/F}$, and $\Lambda_{\rm FM}$ and the relative angle $\phi_{AB}$ of the ground state over $\theta\in[0,2\pi]$ [Fig.~\ref{fig:phase_order}]. With increasing $\theta$, the ground state passes through the AM/N, noncoplanar, F/F, FM, noncoplanar, and AM/N phases, separated at $\theta=0.953$, $1.37$, $\pi$, $4.915$, and $2\pi-0.953$. The AM/N order parameter $\Lambda_{\rm AM/N}$ decreases from unity in the AM/N phase and falls to nearly zero within the noncoplanar phase, whereas $\Lambda_{\rm F/F}$ and $\Lambda_{\rm FM}$ are close to unity throughout the F/F and FM phases, respectively [Fig.~\ref{fig:phase_order}(a)]. The relative angle is $\phi_{AB}=\pi/2$ in the AM/N phase, $\pi$ in the F/F phase, and $0$ in the FM phase, while it is not defined in the noncoplanar phase [Fig.~\ref{fig:phase_order}(b)]. {Throughout the AM/N phase, $\theta<0.953$ and $\theta>2\pi-0.953$, the ground state is the canted AM/N state of Sec.~\ref{sec:canting}.} The map $\theta\to2\pi-\theta$ is exact, because every $J_1$ bond joins $G_A$ and $G_B$ while every $J_4$ bond stays within one set, so reversing all spins of $G_B$ is equivalent to $J_1\to-J_1$; it exchanges $\Lambda_{\rm F/F}$ and $\Lambda_{\rm FM}$, leaves $\Lambda_{\rm AM/N}$ invariant, and maps $\phi_{AB}$ onto $\pi-\phi_{AB}$.

\subsection{Iterative minimization and constrained states}
\label{sec:minimization}
{Ground states are obtained by iterative minimization from random initial spins.} We obtain each stationary state by iterating the damped local-field update
\bea
\bS_i\;\rightarrow\;\frac{(1-\lambda)\bS_i-\lambda\hat{\bm h}_i}{|(1-\lambda)\bS_i-\lambda\hat{\bm h}_i|},\qquad \bm h_i=\sum_jJ_{ij}\bS_j,
\label{eq:update}
\eea
where $\hat{\bm h}_i=\bm h_i/|\bm h_i|$, $J_{ij}=J_1$ ($J_4$) on $J_1$ ($J_4$) bonds, and $\lambda=0.2$ denotes the damping. A fixed point aligns every spin antiparallel to its local field, so the torque $\tau=\max_i|\bS_i\times\bm h_i|$ vanishes and the configuration $\{\bS_i^0\}$ is a stationary point of $E=\frac12\sum_{ij}J_{ij}\bS_i\cdot\bS_j$; we stop at $\tau<10^{-11}$. We follow three constrained states by removing the constrained components of $\bm h_i$ as
\bea
\left\{
\begin{array}{ll}
S_i^y=0,\;\; h_i^y\rightarrow0 & \text{(canted coplanar)},\\[3pt]
S_i^x=S_i^y=0,\;\; h_i^x,h_i^y\rightarrow0,\;\; \bS_i=\pm\hat{\bm z}\ \text{on}\ V_{A/B}\ \text{initially} & \text{(F/F)},\\[3pt]
S_i^x=S_i^y=0,\;\; h_i^x,h_i^y\rightarrow0,\;\; \bS_i=+\hat{\bm z}\ \text{initially} & \text{(FM)},
\end{array}
\right.
\label{eq:constraints}
\eea
where the canted coplanar state starts from the rigid orthogonal N\'eel state, $\bS_i=\epsilon_i\hat{\bm z}$ on $V_A$ and $\epsilon_i\hat{\bm x}$ on $V_B$. Each constrained fixed point is stationary for the full energy, because the local fields of a configuration in the $xz$ plane (along $\hat{\bm z}$) lie in that plane (along $\hat{\bm z}$).

We continue each state in $\theta$ over $0.02\le\theta\le6.28$ in steps of $0.02$, refined to $0.002$ within $0.03$ of each boundary, both forward and backward. Each chain starts from the rigid state with random noise of amplitude $10^{-4}$, for ten noise realizations, and at each $\theta$ we keep the lower-energy state of the two directions. {For $0.8\le\theta\le1.4$, we keep the lowest-energy state among continuation from smaller $\theta$, continuation from larger $\theta$, and random initial spins.}

\subsection{Hessian analysis}
\label{sec:hessian}
{We relax all spins but evaluate the Hessian only for the spins at $r=|\br(i)|<R-6$, where $R$ denotes the patch radius, holding the outer spins at their values in $\{\bS_i^0\}$. Each inner spin is parametrized as}
\bea
\bS_i=\sqrt{1-|\bm a_i|^2}\,\bS_i^0+\sum_{\mu=1,2}a_i^\mu\,\bm e_i^\mu,
\label{eq:tangent}
\eea
where $\bm e_i^1$ and $\bm e_i^2$ denote orthonormal vectors perpendicular to $\bS_i^0$ and $\bm a_i=\sum_\mu a_i^\mu\bm e_i^\mu$ the canting of site $i$. The second derivative of $E$ gives the Hessian matrix
\bea
\mathcal{H}_{i\mu,j\nu}=\frac{\partial^2E}{\partial a_i^\mu\,\partial a_j^\nu}\bigg|_{\bm a=0}=J_{ij}\,\bm e_i^\mu\cdot\bm e_j^\nu+\delta_{ij}\delta_{\mu\nu}\kappa_i,\qquad \kappa_i=-\bS_i^0\cdot\bm h_i^0 ,
\label{eq:hessian}
\eea
with $\bm h_i^0=\sum_jJ_{ij}\bS_j^0$ and $i,j$ restricted to $r<R-6$. {Since the energy is isotropic, a uniform rotation of all spins costs no energy and gives three zero eigenvalues of the unrestricted Hessian. Holding the outer spins removes these zero modes, together with out-of-plane modes localized at the patch edge that become unstable already at $\theta\approx0.89$, so that $\mathcal{H}$ probes the bulk.} Hence, $\{\bS_i^0\}$ is a local minimum if all eigenvalues of $\mathcal{H}$ are positive and a saddle point if one is negative.

\begin{figure}[h]
\centering
\includegraphics[width=\linewidth]{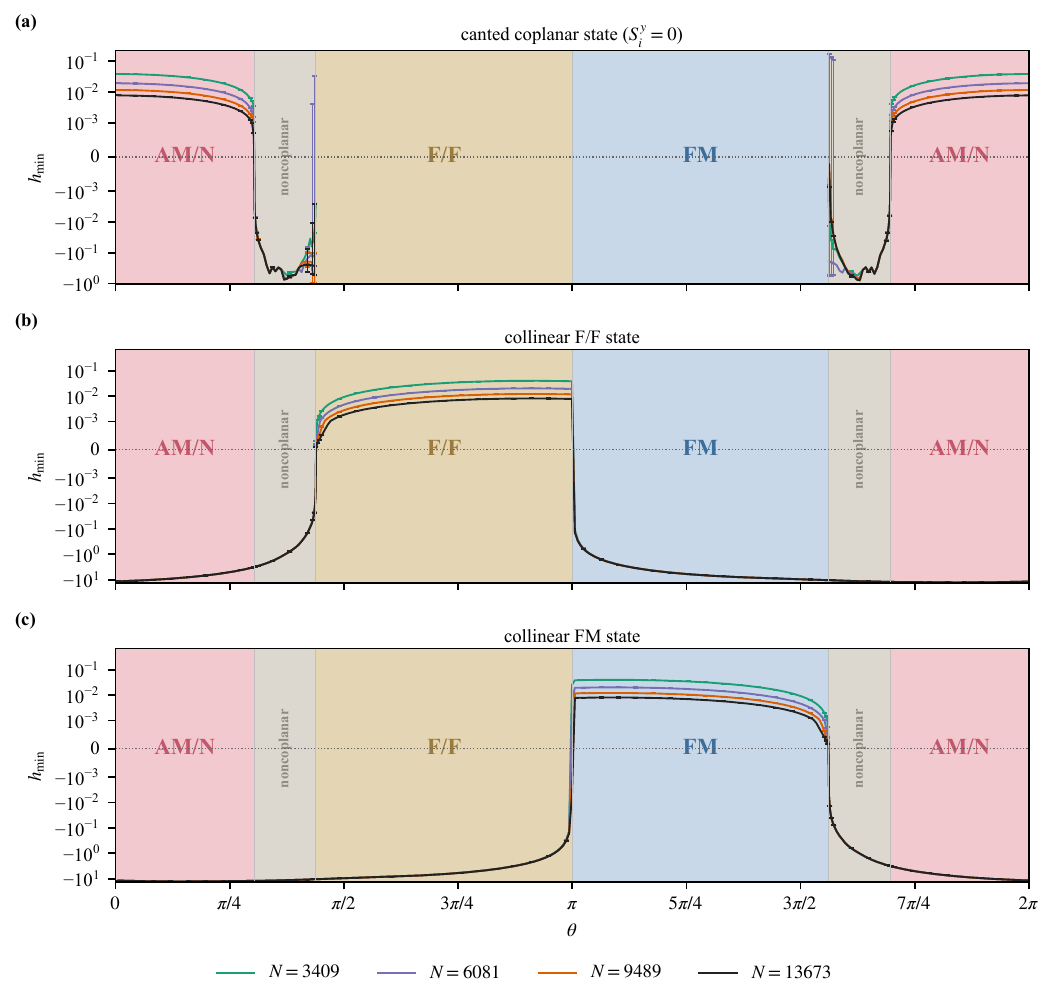}
\caption{Lowest eigenvalue $h_{\min}$ of the Hessian $\mathcal{H}$ [Eq.~\eqref{eq:hessian}], with the spins at $r\ge R-6$ held fixed, versus $\theta\in[0,2\pi]$ of (a) the canted coplanar state with $S_i^y=0$, (b) the collinear F/F state, and (c) the collinear FM state, for $N=3409$, $6081$, $9489$, and $13673$ ($R=30$, $40$, $50$, and $60$). Lines show the mean and bars the standard deviation over ten noise realizations; the vertical axis is linear for $|h_{\min}|<10^{-3}$ and logarithmic beyond, and the dotted line marks $h_{\min}=0$. In (a), the shaded intervals mark the F/F ($1.3789<\theta<\pi$) and FM ($\pi<\theta<2\pi-1.3789$) phases, where the canted coplanar state is not the ground state and $h_{\min}$ is not shown. Vertical lines mark $\theta=0.953$, $1.379$, $\pi$, $2\pi-1.379$, and $2\pi-0.953$. At $N=13673$, $h_{\min}$ changes sign at $\theta=0.9533$ and $5.3299$ in (a), $1.3789$ in (b), and $4.9041$ in (c).}
\label{fig:hessian}
\end{figure}

{The AM/N phase ends at $\theta_c=0.953$, where a bulk out-of-plane rotation of the N\'eel order becomes a zero mode of the Hessian of the canted coplanar state [Fig.~\ref{fig:hessian}(a)]. At the same $\theta$, the second derivative of the ground-state energy jumps, indicating a continuous transition. Beyond $\theta_c$ the spins leave the common plane, realizing the noncoplanar phase. The F/F phase sets in at $\theta\approx1.37$, in agreement within $0.01$ with $\theta=1.379$, where the Hessian of the collinear F/F state acquires a negative eigenvalue [Fig.~\ref{fig:hessian}(b)].}

We denote by $h_{\min}$ the lowest eigenvalue of $\mathcal{H}$, computed on patches with $R=30$, $40$, $50$, and $60$ ($N=3409$, $6081$, $9489$, and $13673$). At $R=60$, $h_{\min}$ of the canted coplanar state changes sign at $\theta=0.9533\approx\theta_c$ and at $5.3299\approx2\pi-\theta_c$, where $\Lambda_{\rm AM/N}$ begins to decrease steeply [Figs.~\ref{fig:phase_order}(a) and \ref{fig:hessian}(a)]. Likewise, $h_{\min}$ of the collinear F/F state changes sign at $\theta=1.3789$, below which $\Lambda_{\rm F/F}$ falls from unity to nearly zero within $\Delta\theta\approx0.03$ [Fig.~\ref{fig:hessian}(b)]. That of the collinear FM state changes sign at $\theta=4.9041$ [Fig.~\ref{fig:hessian}(c)]. {From $N=3409$ to $N=13673$, the zero crossings move from $\theta=0.9536$ to $0.9533$, from $5.3296$ to $5.3299$, from $1.3788$ to $1.3789$, and from $4.9043$ to $4.9041$.} Each zero crossing shifts by less than $4\times10^{-4}$ from $R=30$ to $R=60$ and by less than $10^{-4}$ between noise realizations.

\subsection{Spin structure of the ordered states}
\label{sec:states}
\begin{figure}[h]
\centering
\includegraphics[width=\linewidth]{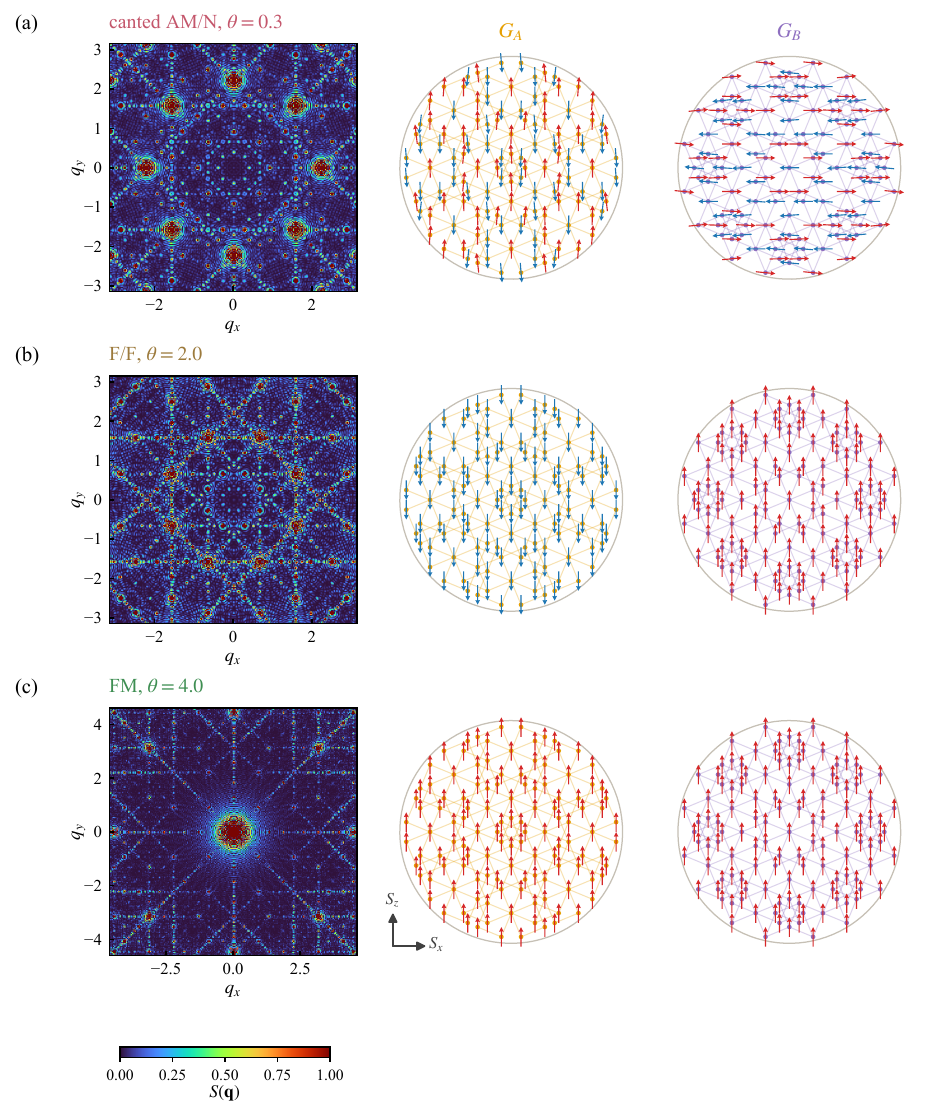}
\caption{Spin structure factor $S(\mathbf q)$ (left) and spins of $G_A$ and $G_B$ within the distance $r=|\br(i)|<8$ from the central vertex (middle and right) of (a) the canted AM/N state at $\theta=0.3$, (b) the F/F state at $\theta=2.0$, and (c) the FM state at $\theta=4.0$. $S(\mathbf q)$ is computed from the $N=13673$ sites with $r<60$ of the $R=100$ patch; the color scale is clipped at $1$. As in Fig.~2 of the main text, the arrows show $(S_x,S_z)$ and are red (blue) for a positive (negative) projection onto the ordering axis of their set; faint lines are $J_4$ bonds. The AM/N state is the relaxed ground state; the F/F and FM states are the exact collinear ground states.}
\label{fig:states}
\end{figure}

We compare the spin structure factor $S(\mathbf q)=N^{-1}|\sum_i\bS_ie^{i\mathbf q\cdot\br(i)}|^2$ and the spins of the three ordered states, the canted AM/N state at $\theta=0.3$, the F/F state at $\theta=2.0$, and the FM state at $\theta=4.0$, where both couplings are ferromagnetic [Fig.~\ref{fig:states}]. Here, $\bS_i$ denotes the spin at site $i$, $\mathbf q$ the wave vector, and $N$ the number of sites in the sum. In the AM/N state, the strongest peaks lie at $|\mathbf q|=\pi/\sqrt2$ along the eightfold axes, half the wave vector $\pi\sqrt2$ of the strongest nonzero-$\mathbf q$ peaks of the FM state [Figs.~\ref{fig:states}(a) and \ref{fig:states}(c)].

\section{Orthogonal locking of the two ordering axes: the canted AM/N state}
\label{sec:canting}

At $\theta=0$, the two sets order independently, leaving the angle $\phi_{AB}$ between their ordering axes arbitrary. Here we show that a small $J_1$ selects $\phi_{AB}=\pi/2$ through a second-order canting of the spins, whose pattern is exactly solvable in the small-$J_1$ limit [Fig.~\ref{fig:canting}]. The resulting state, in which the two ordering axes lock orthogonally and every spin cants within their common plane, retains the spin-group symmetry $[C_{2x}\parallel C_{8z}]$ of the altermagnet. Throughout, $\epsilon_i$ denotes the staggered sign of Eq.~\eqref{eq:neel_sign}.

\subsection{Setting}
\begin{figure}[h]
\centering
\includegraphics[width=\linewidth]{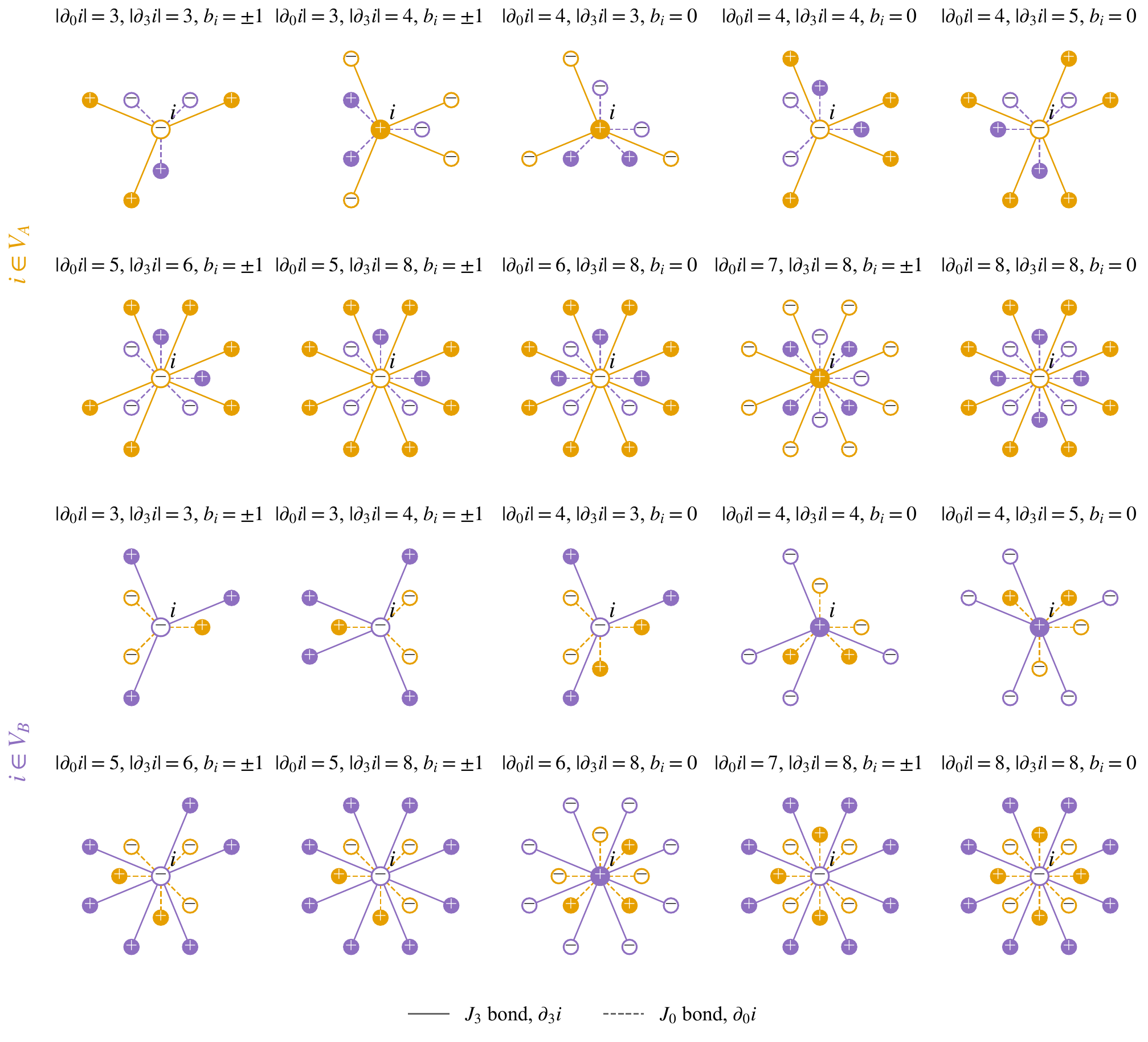}
\caption{The ten local environments $(|\partial_1 i|,|\partial_4 i|)$ of the bulk of the tiling, which occur in both sets, each drawn at a representative site with $b_i\ge0$, $i\in V_A$ (top two rows) and $i\in V_B$ (bottom two rows), with its $J_4$ neighbors $\partial_4 i$ (solid, same set) and $J_1$ neighbors $\partial_1 i$ (dashed, other set); orange marks $V_A$ and purple $V_B$. Filled and open symbols mark the staggered signs $\epsilon=\pm1$ of the solvable point, and $b_i$ is the signed neighbor count of Eq.~\eqref{eq:bi}. Sites with the opposite sign of $b_i$ are the $C_{8z}$ images of those shown, since $C_{8z}$ maps $b_i\to-b_i$ [Eq.~\eqref{eq:S1}].}
\label{fig:nbhd}
\end{figure}

For a site $i$ we write $\partial_4 i$ for its $J_4$ neighbors (all in the same set, $z_i\equiv|\partial_4 i|$) and $\partial_1 i$ for its $J_1$ neighbors (all in the other set). Only ten combinations $(|\partial_1 i|,|\partial_4 i|)$ occur in the bulk of the tiling, with one representative of each shown in Fig.~\ref{fig:nbhd}. The only local quantity that enters below is the signed count of $J_1$ neighbors,
\bea
b_i\equiv\epsilon_i\sum_{j\in\partial_1 i}\epsilon_j ,
\label{eq:bi}
\eea
which does not vanish in general because the quasiperiodic coordination does not balance the two staggered signs around a site (Fig.~\ref{fig:nbhd}; $|b_i|=2$ occurs only at the patch boundary). We use the N\'eel vector $\bm M_\gamma$ and ordering axis $\bm n_\gamma$ of Eq.~\eqref{eq:neel}, with $\cos\phi_{AB}=\bm n_A\cdot\bm n_B$. The rigid N\'eel state is $\bm S_i=\epsilon_i\bm n_{\gamma(i)}$, for which $\bm M_\gamma=N_\gamma\bm n_\gamma$ with $N_\gamma=|V_\gamma|$. Since the Hamiltonian is isotropic, every state is defined up to a global spin rotation, which we fix by $\bm n_A=\hat z$ and $\bm n_B$ in the $xz$ plane; for $\phi_{AB}=\pi/2$, $\bm n_B=\hat x$. Finally, under the site permutation $\mathcal{R}$ of Sec.~\ref{sec:J4_parity}, $C_{8z}$ preserves the staggered sign on $V_B$, reverses it on $V_A$ [Eq.~\eqref{eq:eps_rotation}], and maps the set of $J_1$ bonds onto itself as
\bea
\epsilon_{\mathcal{R}i}=-\epsilon_i\ \ (i\in V_A),\qquad \epsilon_{\mathcal{R}j}=+\epsilon_j\ \ (j\in V_B),\qquad \mathcal{R}\langle i,j\rangle_1=\langle\mathcal{R}i\,\mathcal{R}j\rangle_1 .
\label{eq:sigma}
\eea

\subsection{Perturbative expansion in small cantings}
\emph{Rigid N\'eel state.} Without canting the energy is
\bea
E^{(0)}=-J_4N_4+J_1\cos\phi_{AB}\,S_1,\qquad S_1\equiv\sum_{\langle i,j\rangle_1}\epsilon_i\epsilon_j=\sum_{i\in V_A}b_i ,
\label{eq:E0}
\eea
with $N_4$ the number of $J_4$ bonds. Since $\mathcal{R}$ only permutes the $J_1$ bonds among themselves, the sum of $\epsilon_{\mathcal{R}i}\epsilon_{\mathcal{R}j}$ over all bonds equals $S_1$. By contrast, $\mathcal{R}$ reverses exactly one of the two signs in each term [Eq.~\eqref{eq:sigma}], so
\bea
S_1=\sum_{\langle i,j\rangle_1}\epsilon_{\mathcal{R}i}\,\epsilon_{\mathcal{R}j}=-\sum_{\langle i,j\rangle_1}\epsilon_i\epsilon_j=-S_1\;\Rightarrow\;S_1=0,
\qquad\text{and likewise}\qquad b_{\mathcal{R}i}=-b_i\ \ \text{for every site } i .
\label{eq:S1}
\eea
The term linear in $J_1$ therefore vanishes identically, leaving the rigid N\'eel states degenerate in $\phi_{AB}$.

\emph{Second order in the canting.} We let each spin deviate from its ordering axis in an arbitrary direction perpendicular to it,
\bea
\bm S_i=\epsilon_i\Bigl(\sqrt{1-|\delta\bm n_i|^2}\;\bm n_{\gamma(i)}+\delta\bm n_i\Bigr),\qquad \delta\bm n_i\perp\bm n_{\gamma(i)} .
\label{eq:variation}
\eea
To write the $J_1$ term we decompose each ordering axis along and perpendicular to the other, and each deviation into its component in the plane of $\bm n_A$ and $\bm n_B$ and its component perpendicular to that plane,
\bea
\bm n_B&=&\cos\phi_{AB}\,\bm n_A+\sin\phi_{AB}\,\hat{\bm m},\qquad
\bm n_A=\cos\phi_{AB}\,\bm n_B+\sin\phi_{AB}\,\hat{\bm m}',\nonumber\\
\delta\bm n_i&=&v_i\,\hat{\bm m}+w_i\,\hat{\bm y}\ \ (i\in V_A),\qquad
\delta\bm n_j=v_j\,\hat{\bm m}'+w_j\,\hat{\bm y}\ \ (j\in V_B),
\label{eq:mdef}
\eea
where $\hat{\bm m}\perp\bm n_A$ and $\hat{\bm m}'\perp\bm n_B$ are the unit vectors in the plane of $\bm n_A$ and $\bm n_B$ and $\hat{\bm y}$ is the normal of that plane. Since $\delta\bm n_i\perp\bm n_A$, the $\cos\phi_{AB}$ part drops out of $\delta\bm n_i\cdot\bm n_B=\sin\phi_{AB}\,v_i$, and likewise $\bm n_A\cdot\delta\bm n_j=\sin\phi_{AB}\,v_j$. Expanding to second order in $\delta\bm n$ and first order in $J_1$ (the two are of the same order, $|\delta\bm n|\sim J_1/J_4$, as the solution below confirms),
\bea
E(\phi_{AB},\delta\bm n)&=&E^{(0)}+\frac{J_4}{2}\sum_{\langle i,k\rangle_4}|\delta\bm n_i-\delta\bm n_k|^2
+J_1\sum_{\langle i,j\rangle_1}\epsilon_i\epsilon_j\bigl(\delta\bm n_i\cdot\bm n_B+\bm n_A\cdot\delta\bm n_j\bigr)\nonumber\\
&=&E^{(0)}+\frac{J_4}{2}\sum_{\gamma=A,B}\ \sum_{i,k\in V_\gamma}(L_\gamma)_{ik}\,\delta\bm n_i\cdot\delta\bm n_k
+J_1\sin\phi_{AB}\Bigl(\sum_{i\in V_A}b_i\,v_i+\sum_{j\in V_B}b_j\,v_j\Bigr)\nonumber\\
&=&E^{(0)}+\sum_{\gamma=A,B}\Bigl[\frac{J_4}{2}\,v_\gamma^{\!\top}L_\gamma\,v_\gamma+J_1\sin\phi_{AB}\;b_\gamma^{\!\top}v_\gamma\Bigr]
+\frac{J_4}{2}\sum_{\gamma=A,B}w_\gamma^{\!\top}L_\gamma\,w_\gamma ,
\label{eq:E2}
\eea
where $L_\gamma$ denotes the graph Laplacian of the $J_4$ graph $G_\gamma$,
\bea
(L_\gamma)_{ik}=\begin{cases}z_i, & i=k,\\ -1, & \langle i,k\rangle_4,\\ 0, & \text{otherwise},\end{cases}
\label{eq:laplacian}
\eea
with $v_\gamma$, $w_\gamma$, and $b_\gamma$ the column vectors of the $v_i$, $w_i$, and $b_i$ over $i\in V_\gamma$ and $^{\!\top}$ the transpose.

The $J_4$ term in the first line of Eq.~\eqref{eq:E2} follows from Eq.~\eqref{eq:variation} with $\epsilon_i\epsilon_k=-1$ on every $J_4$ bond: a bond costs energy only if its two spins tilt differently. Expanding $|\delta\bm n_i-\delta\bm n_k|^2=|\delta\bm n_i|^2+|\delta\bm n_k|^2-2\,\delta\bm n_i\cdot\delta\bm n_k$, the term $|\delta\bm n_i|^2$ appears once for each of the $z_i$ bonds of site $i$ and the cross term once per bond, which gives the quadratic form of the second line with the Laplacian of Eq.~\eqref{eq:laplacian}. Collecting the $J_1$ bond sum site by site produces $b_i$. The third line uses Eq.~\eqref{eq:mdef}: the in-plane components $v_i$ and the out-of-plane components $w_i$ separate, and only the former couple to $J_1$. Only the component of the exchange field transverse to a spin's own axis can tilt a fixed-length spin, so the $J_1$ term is proportional to $\sin\phi_{AB}$ and vanishes identically for collinear N\'eel vectors. Since the $w_i$ enter only through the positive cost, they vanish at the minimum: the state is coplanar, every spin tilting within the plane spanned by the two ordering axes [Fig.~\ref{fig:canting}(a),(b)].

\emph{Solution.} With the $w_i$, not coupled to $J_1$, vanishing separately, the stationarity condition $\partial E/\partial v_i=0$ of the third line of Eq.~\eqref{eq:E2}, where the $J_4$ term gives $J_4(L_\gamma v_\gamma)_i$ and the $J_1$ term $J_1\sin\phi_{AB}\,b_i$, is one linear equation per site,
\bea
J_4\Bigl(z_iv_i-\sum_{k\in\partial_4 i}v_k\Bigr)=-J_1\sin\phi_{AB}\,b_i ,\qquad\text{i.e.}\qquad J_4L_\gamma v_\gamma=-J_1\sin\phi_{AB}\,b_\gamma .
\label{eq:linear}
\eea
Two properties of $L_\gamma$ follow from the fact that every row of it sums to zero, $z_i-\sum_{k\in\partial_4 i}1=0$. First, tilting all spins of a set by the same amount, $v_i=c$ for every $i\in V_\gamma$, costs no $J_4$ energy, $L_\gamma\mathbf 1=0$ with $\mathbf 1=(1,\dots,1)$. This uniform tilt is a rigid rotation of the ordering axis $\bm n_\gamma$, i.e.\ a change of $\phi_{AB}$, which we treat as a separate parameter. Second, summing Eq.~\eqref{eq:linear} over all sites of the set gives $\mathbf 1^{\!\top}L_\gamma v_\gamma=0$, so the right-hand side must sum to zero as well, $\sum_{i\in V_\gamma}b_i=S_1=0$. The canting equations therefore have a solution precisely because of Eq.~\eqref{eq:S1}, which also ensures that the transverse field does not drive $\phi_{AB}$ at first order. The solution is determined up to a constant added to all $v_i$; we fix it by $\sum_{i\in V_\gamma}v_i=0$, which keeps the axis $\bm n_\gamma$ of the expansion exactly parallel to the N\'eel vector $\bm M_\gamma$ of Eq.~\eqref{eq:neel} also in the canted state.

Since the right-hand side of Eq.~\eqref{eq:linear} is proportional to $J_1\sin\phi_{AB}/J_4$, so is the solution, $v_\gamma=(J_1\sin\phi_{AB}/J_4)\,u_\gamma$. Here, the dimensionless pattern $u_\gamma$ solves $L_\gamma u_\gamma=-b_\gamma$ with $\sum_iu_i=0$ and depends only on the geometry of the patch, with $u_i\approx-b_i/z_i$ if the tilts of the neighbors are neglected. Inserting the solution into Eq.~\eqref{eq:E2}, where the linear term equals minus twice the quadratic one, gives
\bea
E(\phi_{AB})\equiv\min_{\delta\bm n}E(\phi_{AB},\delta\bm n)=E^{(0)}-\frac{J_1^2}{2J_4}\,c_2N\sin^2\phi_{AB},
\qquad
c_2\equiv\frac1N\sum_{\gamma=A,B}u_\gamma^{\!\top}L_\gamma\,u_\gamma=-\frac1N\sum_{i}b_iu_i .
\label{eq:solution}
\eea
On the $R=30$ patch, $c_2=0.272$ (the local estimate $\sum_ib_i^2/z_i/N=0.205$ is smaller because neighboring spins tilt together and share the $J_4$ cost). Since $c_2>0$, the collinear states $\phi_{AB}=0,\pi$ are saddle points of the energy and $\phi_{AB}=\pi/2$ is the minimum [Fig.~\ref{fig:canting}(d)]. The Hessian of Eq.~\eqref{eq:hessian}, evaluated on the whole patch with no spins held fixed, confirms this at $\theta=0.1$, $0.3$, $0.6$ and $0.85$, with a negative eigenvalue at $\phi_{AB}=0,\pi$ and only positive eigenvalues at the relaxed canted state apart from the three zero modes of a global spin rotation. The relation $E(\phi_{AB})=E(\pi-\phi_{AB})$ holds to all orders because $C_{8z}$ acts as $\bm n_A\to-\bm n_A$, so only even powers of $\cos\phi_{AB}$ appear. The energy gain is quadratic in $J_1$ and therefore independent of its sign.

\subsection{Symmetry of the canted state}
\begin{figure}[h]
\centering
\includegraphics[width=\linewidth]{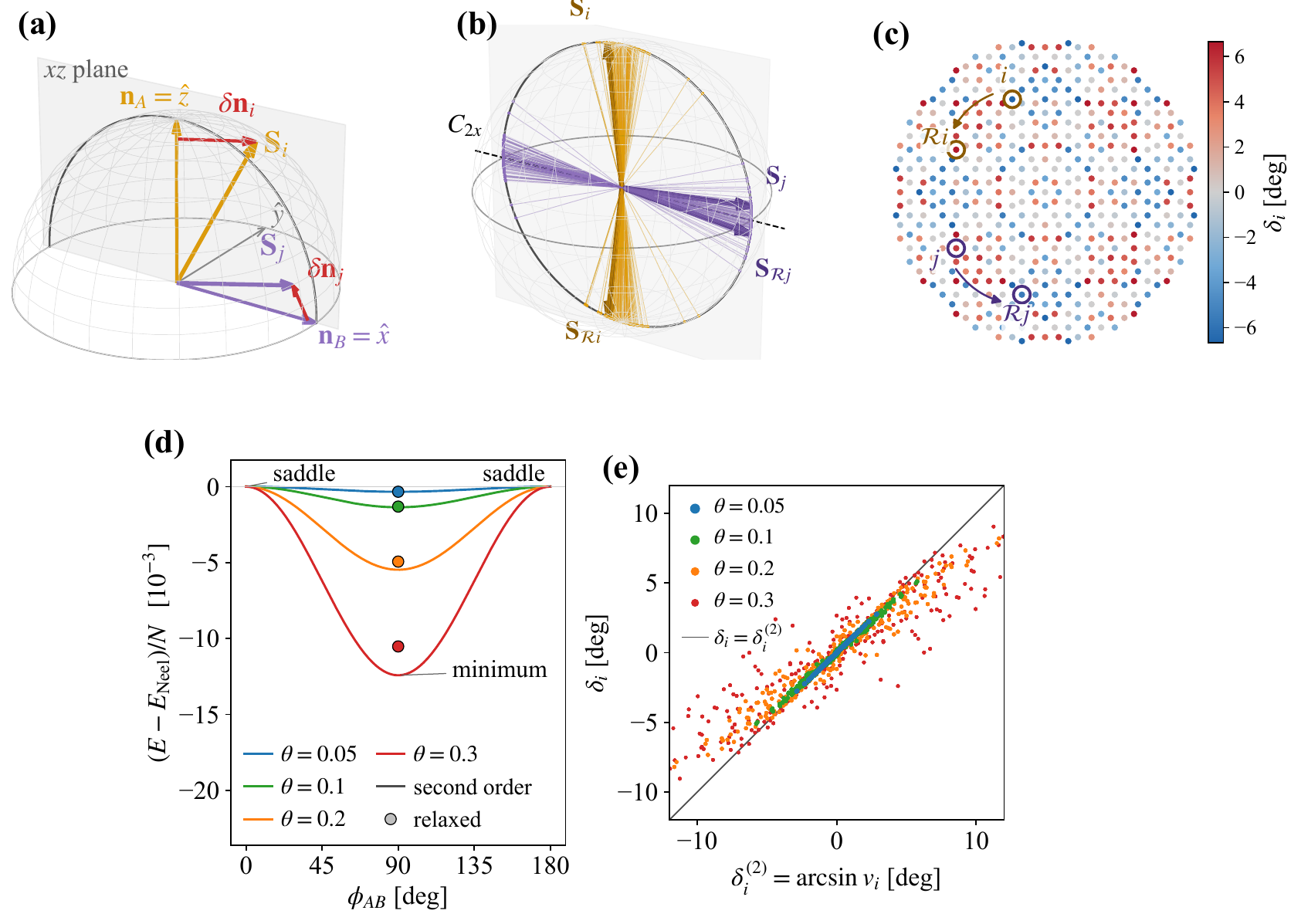}
\caption{Orthogonal locking of the two ordering axes by second-order canting ($R=30$, $N=3409$). (a) The ordering axes $\bm n_A=\hat z$, $\bm n_B=\hat x$ on the upper half of the unit sphere in spin space, whose points are the spin directions; the gray rectangle is the $xz$ plane spanned by them. A $V_A$ spin $\bm S_i$ and a $V_B$ spin $\bm S_j$ deviate from their axes by the tangent vectors $\delta\bm n_i\perp\bm n_A$ and $\delta\bm n_j\perp\bm n_B$ (red) of Eq.~\eqref{eq:variation}; at the minimum the components of $\delta\bm n$ perpendicular to the $xz$ plane vanish. (b) All $N$ spins of the relaxed ground state at $\theta=0.3$, drawn from the origin onto the unit sphere in spin space as translucent lines (orange $V_A$, purple $V_B$): each set fans out from $\pm\bm n_A$ and $\pm\bm n_B$ within the $xz$ plane, and no spin leaves it. The opaque arrows are two $C_{8z}$ pairs, $(\bm S_i,\bm S_{\mathcal R i})$ with $i\in V_A$ and $(\bm S_j,\bm S_{\mathcal R j})$ with $j\in V_B$; each pair is mapped onto itself by the $\pi$ rotation $C_{2x}$ about the dashed $x$ axis. (c) The same state in real space, colored by the canting angle $\delta_i$ of each site ($r<11.5$), i.e.\ the signed angle by which $\epsilon_i\bm S_i$ turns from its own ordering axis towards the other one within the $xz$ plane ($\delta_i\simeq\arcsin v_i$ at small $J_1$). The rotation $C_{8z}$ about the center of the patch reverses the canting angle, $\delta_{\mathcal R i}=-\delta_i$, as shown for the marked sites of (b). Since $\delta_i$ is blind to the overall spin sign, (c) fixes only the real-space part; the spin-space part $C_{2x}$ (not $C_{2z}$) is read from the full spins in (b). (d) Energy of the relative angle for $\theta=0.05$, $0.1$, $0.2$ and $0.3$: second-order result of Eq.~\eqref{eq:solution} (lines, $c_2=0.272$) and relaxed ground state of the full model (circles, placed at its own $\phi_{AB}$ and energy). (e) Canting angle $\delta_i$ of every bulk site ($r<24$) of the relaxed ground state against the second-order prediction $\delta_i^{(2)}=\arcsin v_i$ of Eq.~\eqref{eq:linear}, for $\theta=0.05$, $0.1$, $0.2$ and $0.3$ (the smaller $\theta$ is drawn on top); the line marks $\delta_i=\delta_i^{(2)}$.}
\label{fig:canting}
\end{figure}

Because $b_{\mathcal{R}i}=-b_i$ [Eq.~\eqref{eq:S1}] and $L_\gamma$ commutes with $\mathcal{R}$, the canting obeys $v_{\mathcal{R}i}=-v_i$. With $\bm n_A=\hat z$, $\bm n_B=\hat x$ this gives, for $i\in V_A$, $\bm S_{\mathcal{R}i}=\epsilon_{\mathcal{R}i}(\hat z+v_{\mathcal{R}i}\hat x)=\epsilon_i(-\hat z+v_i\hat x)=C_{2x}\bm S_i$, and for $j\in V_B$, $\bm S_{\mathcal{R}j}=\epsilon_j(\hat x-v_j\hat z)=C_{2x}\bm S_j$, where $C_{2x}$ is the $\pi$ rotation of the spins about the ordering axis $\bm n_B=\hat x$ of $G_B$. The spin-group operation $[C_{2x}\parallel C_{8z}]$ that relates the two spin channels of the transmission (Sec.~\ref{sec:transport}) is therefore an exact symmetry of the canted ground state, not only of the rigid N\'eel spin texture [Fig.~\ref{fig:canting}(b),(c)]. Conversely, this symmetry fixes the relative angle at $\phi_{AB}=\pi/2$, independently of the size of the canting. Applied to the N\'eel vectors of Eq.~\eqref{eq:neel}, it requires $\bm M_A=-C_{2x}\bm M_A$ and $\bm M_B=C_{2x}\bm M_B$, so $\bm M_A\perp\hat x$ and $\bm M_B\parallel\hat x$. We refer to this symmetry-protected state, orthogonal N\'eel vectors with a coplanar canting, as the canted AM/N state, the AM/N ground state throughout the red region of Fig.~2 of the main text.

We compare the site-by-site tilt $v_i$ of Eq.~\eqref{eq:linear} with the ground state of the full model obtained by iterative minimization from random spins [Fig.~\ref{fig:canting}(e)]. At $\theta=0.05$ ($0.3$), the correlation is $0.9998$ ($0.93$) and the slope $0.97$ ($0.63$), while the energy gain agrees to $1\%$ ($15\%$), with the intermediate values $\theta=0.1$ and $0.2$ interpolating smoothly. The relaxed angle is $\phi_{AB}=90.0^\circ$ at every $\theta$, and the spin-group residual $\max_i|\bm S_{\mathcal{R}i}-C_{2x}\bm S_i|$ and the out-of-plane component $\max_i|S_i^y|$ stay at the $10^{-6}$ level of the convergence tolerance.

\section{Transport setup}
\label{sec:transport}

\subsection{Model and contacts}
\label{sec:model_en}
The transport calculations use the Hamiltonian $H_{\rm el}$ of Eq.~(8) of the main text on a finite Ammann--Beenker patch of $N$ sites at $(x_i,y_i)$, with the spins $\bm m_i$ of the relaxed ground state. Lengths are in units of the $J_1$ bond length and energies in units of the hopping, $t=1$, so all energies are dimensionless. Including spin, $H_{\rm el}$ is a sparse $2N\times2N$ matrix whose two spin blocks share the hopping and are coupled only through the on-site exchange term. The leads attach to the boundary set $\mathcal{B}$, obtained from the Delaunay triangulation of the patch by keeping the triangles of circumradius at most $1.2$ and collecting the sites on edges that belong to a single kept triangle, so that $\mathcal{B}$ follows the jagged edge of the patch rather than its convex hull.

\subsection{Lead model and surface Green's function}
\label{sec:lead_en}
Each lead is a semi-infinite square lattice with hopping $t_{\rm lead}=t$, whose principal layer is a chain of $M_{\rm lead}$ sites,
\bea
h^{\rm lead}=t_{\rm lead}\sum_{m=1}^{M_{\rm lead}-1}\bigl(\ket{m}\bra{m+1}+\ket{m+1}\bra{m}\bigr),
\label{eq:h_lead}
\eea
with neighboring layers coupled by $\tau_{\rm lead}=t_{\rm lead}I_{M_{\rm lead}}$. The leads are nonmagnetic, so one spin-independent surface Green's function $g^{\rm lead}(E)$ serves both leads and both spin components. It solves the self-consistent Dyson equation
\bea
g^{{\rm lead},(n+1)}(E)=\Bigl[(E+i\eta_1)I_{M_{\rm lead}}-h^{\rm lead}-\tau_{\rm lead}\,g^{{\rm lead},(n)}(E)\,\tau_{\rm lead}^\dagger\Bigr]^{-1},
\label{eq:g_surface_sc_en}
\eea
with $\eta_1=10^{-3}$, iterated from $g^{{\rm lead},(0)}=0$ until the relative change drops below $10^{-6}$.

\subsection{Angle-dependent contacts}
\label{sec:angle_en}
Each boundary site is assigned the polar angle $\phi_i=\mathrm{mod}[\mathrm{atan2}(y_i,x_i),2\pi]$. For a rotation angle $\phi$, the drain $D$ contacts the sites of $\mathcal{B}$ with $\phi_i\in[\phi,\phi+\pi/4)$ and the source $S$ those with $\phi_i\in[\phi+\pi,\phi+5\pi/4)$, with intervals taken modulo $2\pi$ and the sites of each sector ordered by angle. The number of contacts at $\phi=0$ defines $M_{\rm lead}$ ($21$ at $R=29$), which is kept fixed for all $\phi$. When the two sectors select unequal numbers of sites, each site $(x_i,y_i)$ of the sector with $M_{\rm lead}$ sites is paired with the site nearest to $(-x_i,-y_i)$, and angles at which neither sector has $M_{\rm lead}$ sites are skipped. The ordered contact list $\mathcal{C}_\nu(\phi)=\{i_1,\dots,i_{M_{\rm lead}}\}\subset\mathcal{B}$ of lead $\nu\in\{S,D\}$ defines the $N\times M_{\rm lead}$ connection matrix
\bea
[W_\nu(\phi)]_{ik}=\delta_{i,\,i_k},
\label{eq:W}
\eea
which couples the $k$th site of the principal layer to the contact $i_k$ with amplitude $t_c=t$. The lead self-energies are
\bea
\Sigma_\nu(E,\phi)=t_c^2\,W_\nu(\phi)\,g^{\rm lead}(E)\,W_\nu^\dagger(\phi)\otimes \sigma_0,\qquad \nu=S,D,
\label{eq:Sigma_lead}
\eea
where $\sigma_0$ denotes the identity in spin space. All $\phi$ dependence thus enters through the contact lists $\mathcal{C}_\nu(\phi)$.

\subsection{Transmission}
\label{sec:NEGF_en2}
The self-energies enter the retarded Green's function $G=[E+i0^+-H_{\rm el}-\Sigma_S-\Sigma_D]^{-1}$ of the main text, with $0^+$ implemented as $10^{-6}$, and the broadening matrices $\Gamma_\nu=i(\Sigma_\nu-\Sigma_\nu^\dagger)$. The source injects electrons with spin $\pm\hat a$ through $\Gamma_S^{\pm\hat a}=\Gamma_S\,(I_N\otimes\mathcal{P}_{\pm\hat a})$, where $\mathcal{P}_{\pm\hat a}=(\sigma_0\pm\hat a\cdot\bm{\sigma})/2$ denotes the spin projector of the main text, and the drain collects both spin components, which gives $T_{\pm\hat a}(E,\phi)$ of Eq.~(10) of the main text. The angle $\phi$ is measured from $+\hat x$ counterclockwise and sampled in steps of $1^\circ$ over $[0,2\pi)$ for Fig.~4 of the main text and of $3^\circ$ over $[0,\pi)$ for Fig.~\ref{fig:theta_transport}.
\subsection{Spin polarization along the ordering axis of \texorpdfstring{$G_B$}{GB}}
\label{sec:transport_x}
\begin{figure}[h]
\centering
\includegraphics[width=\linewidth]{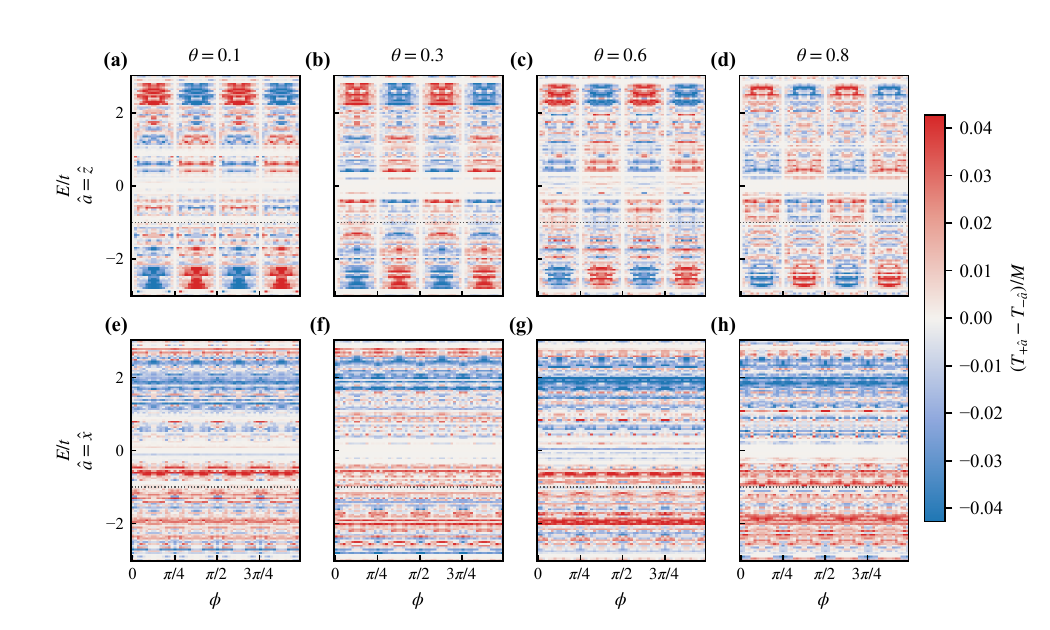}
\caption{Spin-resolved transmission differences per lead channel, $(T_{+\hat z}-T_{-\hat z})/M_{\rm lead}$ (top) and $(T_{+\hat x}-T_{-\hat x})/M_{\rm lead}$ (bottom) on the $(E,\phi)$ plane for the canted AM/N ground state at several $\theta$, $h/t=1$.}
\label{fig:theta_transport}
\end{figure}

Figure~\ref{fig:theta_transport} shows the spin-resolved transmission differences $\Delta T_{\hat a}\equiv T_{+\hat a}-T_{-\hat a}$ on the $(E,\phi)$ plane for the canted AM/N ground state at several $\theta$.
Along $\hat z$, $\Delta T_{\hat z}$ changes sign under $\phi\to\phi+\pi/4$ as required by Eq.~(11) of the main text, whereas each $T_{\pm\hat x}$ is invariant under the same rotation to within $4\times10^{-7}$ for $\theta\ge0.3$ ($5\times10^{-5}$ at $\theta=0.1$), so that $\Delta T_{\hat x}$ is even under $C_{8z}$ and $[C_{2x}\parallel C_{8z}]$ fixes neither its sign nor its magnitude.
A nonzero $\Delta T_{\hat x}$ is nevertheless allowed, since the confined states on $G_B$ near $E=-h$ are polarized along $\pm\hat x$ and leave the residual polarization that splits the $\hat x$ channels of the density of states in Fig.~3(b) of the main text.
This polarization resides on the confined states in the interior of the cluster and is weak at the boundary sites contacted by the leads, whereas $\Delta T_{\hat x}$, about $10\%$ of $T_{\pm\hat x}$ at $\theta=0.3$, changes sign as $\phi$ moves the contacts ($8$, $2$ and $12$ times in $0\le\phi<\pi/4$ for $R=29$, $40$, $57$).
The symmetry thus allows $\Delta T_{\hat x}$, while its sign and magnitude depend on the lead position and on $R$, rather than directly measuring the polarization at the contacted sites.

